\documentclass[11pt]{report}

\usepackage{epsfig}
\usepackage{tabularx,layout,jtb,vmargin,times}
\usepackage{amssymb,amsmath}

\setpapersize{USlegal}
\setmargins{2in}{1.3in}{388pt}{585pt}{12pt}{25pt}{12pt}{35pt}

    \author{Homayoun Bagheri-Chaichian}

    \author{Juozas R. Vaisnys}

   \author{G$\ddot{\mbox{u}}$nter P. Wagner}

\newenvironment{changemargin}[2]{%
 \begin{list}{}{%
  \setlength{\topsep}{0pt}%
 \setlength{\leftmargin}{#1}%
  \setlength{\rightmargin}{#2}%
  \setlength{\listparindent}{\parindent}%
  \setlength{\itemindent}{\parindent}%
  \setlength{\parsep}{\parskip}%
 }%
\item[]}{\end{list}}

    \newcommand{\ed}{\end{document}}
    \newcommand{\be}{\begin{equation} }
    \newcommand{\cij}{C_{i}^{J}}
    \newcommand{\de}{\delta}
    \newcommand{\nsum}{\sum_{i=1}^{n}}
   \newcommand{\n}{\displaystyle}
   \newcommand{\ee}{\end{equation} }
   \newcommand{\la}{\label}
   \newcommand{\limit}[2]{\underset{#1\rightarrow #2}{\lim}}
   \newcommand{\0}{$}
   \newcommand{\nin}{\noindent}

   \newcommand{\spoo}{ {\bf P_{1,1} } }
   \newcommand{\spuJ}{ {\bf  P_{u,J} } }
   \newcommand{\spuDJ}{ {\bf  P_{uD,J} } }
   \newcommand{\spuVJ}{ {\bf  P_{u,V_{max},J} } }
   \newcommand{\spuDVJ}{ {\bf  P_{uD,V_{max},J} } }

   \newcommand{\every}{\forall}
   \newcommand{\spart}[2]{\frac{\partial {#1}}{\partial{#2}}}
   \newcommand{\dopart}[3]{\frac{\partial^{2} {#1}}{\partial{#2}\partial{#3}} }
   \newcommand{\dpj}{\frac{\partial^{2} J}{\partial{E_{1}}\,\partial{E_{2}}}}
   \newcommand{\disspart}[2]{\frac{\de {#1}}{\de{#2}}}
   
   \newcommand{\ifif}{\Leftrightarrow}
   \newcommand{\fset}[2]{{\mathbb{R}^{#1}_{\geq0}\mapsto\mathbb{R}^{#2}}}
   \newcommand{\andd}{~\wedge~}
   \newcommand{\dc}[2]{\stackrel{{#1}{#2}}{\longmapsto}}

    \newcommand{\ddc}[2]
    {
    \overset{{#1}{#2}}
         {
          \underset{\Delta}{\longmapsto}
         }
    }
    \newcommand{\real}[1]{\mathbb{R}^{#1}}
    
\newcommand{\realgg}[1]{\mathbb{R}^{#1}_{>0}}
    \newcommand{\A}{(``\mbox{Stoichiometry: two enz. seq.}")\lfloor p \rfloor}
    \newcommand{\B}{(``\mbox{Const. u \0\rightarrow\0 Unique }\langle u,E_1,E_2,J^*\rangle")\lfloor p \rfloor}
    \newcommand{\C}{(`` \mbox{computable: \0 \langle E_1,E_2,u \rangle \mapsto J~ \0}")\lfloor p \rfloor}
    \newcommand{\D}{(``\mbox{Diffusion Limited \em I}")\lfloor p \rfloor}
    \newcommand{\E}{(``\mbox{Saturable Enzymes}")\lfloor p \rfloor}

    \newcommand{\err}{\varepsilon}
    
    \newcommand{\bea}{\begin{array}{l}}
    \newcommand{\ena}{\end{array}}
    \newcommand{\equ}[1]{(\ref{#1})}

    \newcommand{\ctwo}{C^{J}_1 + C^{J}_2 }

\begin{document}

\begin{center}
\Large \bf The control of phenotype: connecting enzyme variation to physiology \\
\large \rm
\ \\ \ \\ \ \\ \ \\

 Homayoun Bagheri-Chaichian*\dag\ddag\\
Joachim Hermisson* \\
Juozas R. Vaisnys\S \\
 G\0\ddot{\rm u}\0nter P. Wagner* \\ \
\\ 
* Dept. of Ecology and Evolutionary Biology, Yale University, New Haven, CT 06520-8106, U.S.A. \\
\dag  Santa Fe Institute, 1399 Hyde Park Rd., Santa Fe, NM 87501, U.S.A. \\
\S Dept. of Electrical Engineering, Yale University, New Haven, CT 06520, U.S.A. \\
\ddag To whom correspondence should be addressed. Present address: bagheri@santafe.edu
\end{center}

\pagebreak


\pagebreak

\newcommand{\quantone}{(\forall E_1,E_2 \in \real{1}_{\geq
0})}
 \newcommand{\nsumtwo}{\sum_{i=1}^{2}}
\newcommand{\alphamove}{\left(\, \alpha = \frac{\de E_1}{E_1}=\frac{\de E_2}{E_2} \,\right)}
\newcommand{\nakedalphamove}{\left(\, \alpha = \frac{\de E_1}{E_1}=\frac{\de E_2}{E_2} \,\right)}
\newcommand{\alldelta}{\de E_1,\,\de E_2 \in \realgg{1}}

\begin{center}
\section*{Abstract}
\end{center}
\nin Metabolic control analysis (Kacser \& Burns (1973). \it Symp.
Soc. Exp. Biol. \rm \bf 27, \rm 65-104; Heinrich \& Rapoport (1974).
\it Eur. J. Biochem. \rm \bf 42, \rm 89-95) was developed for the
understanding of multi-enzyme networks. At the core of this approach
is the flux summation theorem. This theorem implies that there is an
invariant relationship between the control coefficients of enzymes in
a pathway. One of the main conclusions that has been derived from the
summation theorem is that phenotypic robustness to mutation ( e.g.
dominance ) is an inherent property of metabolic systems and hence
does not require an evolutionary explanation (Kacser \& Burns (1981).
\it Genetics. \rm \bf 97, \rm 639-666; Porteous (1996). \it J. theor.
Biol. \rm \bf 182, \rm 223-232 ). Here we show that for mutations
involving discrete changes (~of any magnitude~) in enzyme
concentration the flux summation theorem does not hold. The scenarios
we examine are two-enzyme pathways with a diffusion barrier, two
enzyme pathways that allow for enzyme saturation and two enzyme
pathways that have both saturable enzymes and a diffusion barrier. Our
results are extendable to sequential pathways with any number of
enzymes. The fact that the flux summation theorem cannot hold in
sequential pathways casts serious doubts on the claim that robustness
with respect to mutations is an inherent property of metabolic
systems.

\pagebreak \section*{1.~Introduction}
How do
changes in enzyme properties alter the physiological
phenotype?
Such knowledge is a key component in understanding the relation between genotype and phenotype.
The two main approaches to this problem have developed into fields
known as metabolic control analysis ( MCA )~\cite{Kacser1973,Heinrich1974,Kacser1995a} and
biochemical systems theory ( BST ) ~\cite{Savageau1976,Savageau1989}.
Results from MCA have had extensive influence in
biochemistry, genetics and evolution ( for an overview see the
entire issue of \em J. theor. Biol.\rm,  182:3 , 1996 ).
The cornerstone of the MCA approach is a theoretical result referred to as the flux summation theorem.
The biological interpretation of the summation theorem is that there are systemic constraints inherent in metabolic
pathways; these constraints limit the magnitude of the effects that changes in enzyme activity can have on flux
through a pathway.
The existence of such constraints has two implications.
In the first place, the summation theorem implies that in general the control of flux in a pathway is shared between enzymes.
Hence rate limiting enzymes are rare~\cite{Kacser1995b}.
 Secondly, the theorem implies
that on the average, mutations that change enzyme concentrations will have a small effect on flux.
This implication has important consequences in evolutionary theory and genetics. It means that phenomena that fall
under the rubric of phenotypic robustness to mutations
( such as
selective neutrality, canalization  and dominance ) are inherent properties of metabolic systems and are not a result
of evolution.
This MCA argument was originally made in reference to the case of dominance, stating that dominance
is an inherent property of metabolic pathways and not a result of evolution \cite{Kacser1981,Keightley1996a,Porteous1996}.

 At the core of the MCA approach is a measure of the
 control exerted by
   each enzyme on flux through a pathway.
   The control coefficient \0 \cij \0
measures how important each enzyme is in its ability to
   affect steady-state flux via changes in enzyme concentration.
   In its original formulation \cite{Kacser1973} the control coefficient was defined as
           \be
   \cij =      {\frac{\delta J}{J}
       }/
       {\frac{\delta E_{i}}{E_{i} }
       },
   \la{cij} \end{equation}
where \0 J \0 is the steady state flux of metabolites through the
pathway
   (net rate of product formation at the end of a pathway), \0 E_i \0 is the concentration of enzyme \0
   i\0,  \0 \delta E_i \0 is a discrete change in enzyme
   concentration and \0 \delta J \0 is the resultant change in
   flux. Hence \0 \cij \0 is basically a ratio of
   percentage change in a pathway's steady state flux to the
   corresponding percentage change in enzyme concentration.

Theoretical results that contradict or object to different conclusions derived
within MCA have been posed several times
 \cite{Cornish-Bowden1987,Giersch1988,Savageau1989,Savageau1992,Grossniklaus1996,Kholodenko1998}.
In each case the possible
misrepresentation of non-linearities in MCA has appeared in a
different guise.
In the first place, Cornish-Bowden~(1987)
showed that in a sequential pathway if the maximal rate \0V_{max}\0
of consecutive enzymes are sequentially decreasing then dominance is
not a necessary property of the pathway.
Hence the possibility exists that dominance can evolve. This
objection was rejected in MCA
 on grounds that the specific arrangement of kinetic
values suggested is a special case that is very unlikely to occur by
chance \cite{Kacser1987}. A second set of objections was then put forth
by Savageau and Sorribas (1989, 1992). They argued that in pathways exhibiting
non-linear behavior that arises from properties such as enzyme-enzyme
interactions, feedback loops or non-sequential pathway structure it
can be shown that dominance is not inherent to the
pathway. Based on
disagreements on the mathematical models used, this second objection
was rejected by MCA proponents \cite{Kacser1991,Kacser1995b}. Nevertheless
   more recent
work indirectly corroborates this objection
\cite{Grossniklaus1996,Kholodenko1998}.

The present paper deals with problems involving the biological
in-applicability of MCA precepts. We show that for discrete changes (
of any magnitude ) in enzyme concentration, the flux summation theorem
does not hold. This means that contrary to the MCA allegations, there
are no a-priori constraints that would require the  magnitude of
mutational effects to be small. Furthermore these results re-open the
possibility that phenotypic robustness to mutations ( e.g. dominance )
is not an inherent property of metabolic pathways and can be a result
of evolution.

Our conclusions do not necessarily imply that the
summation theorem is inherently false as an isolated mathematical
proposition. Rather that it has been applied to the wrong context.
This erroneous application arises due to two main disparities. In the
first place, the argument for the summation theorem in the original
Kacser and Burns paper (1973) is derived using a discrete formulation.
However, in most subsequent work ( e.g. Kacser \& Burns 1981 ) a
continous formulation of the theorem is used. Hence the discrete proof
for the summation theorem does not correspond to the continous form in
which it is frequently used. In the second place, mutations
necessarily involve discrete changes in enzyme activity. Hence,
whether a continous version of the summation theorem holds or not (
which it would in cases where the flux function is homogenous, see
 Giersch 1988 \nocite{Giersch1988} ) does not say anything about
whether a pathway will be robust or exhibit dominance with respect to
discrete changes. In order for a pathway to be robust with respect
to discrete changes, the discrete version of the summation theorem
has to hold. In this paper we show that for dicrete changes of any
size the discrete version does not hold, and hence that robustness
is not an inherent property of a metabolic pathway.

The matter addressed here has been a subject of debate for thirty years. In order to be clear about the objects that we treat and the mathematical properites that pertain to them a rigorous approach to the problem was necessary. The main assumptions and implications we derived are summarized as
follows: \ \\


\subsubsection*{\rm \it INITIAL ASSUMPTION}
\begin{description}
 \item[Assumption 0.] The metabolic pathways studied are two-enzyme pathways. They have
 one input and one output and can reach a steady state flux.
 \end{description}
 \ \\

\subsubsection*{\rm \it CASE A :  DIFFUSION LIMITED PATHWAYS}
 \begin{description}
 \item[Assumption 1.] Maximal rate of input into the metabolic pathway is limited
 by a diffusion process.
\item[Implication 1.] The flux summation theorem does not hold.
 \end{description}
\ \\
\subsubsection*{\rm \it CASE B : SEQUENTIAL PATHWAYS THAT ALLOW FOR
ENZYME SATURATION}
 \begin{description}
 \item[Assumption 1.] The two enzymes are near saturation.
\item[Implication 1.] The flux summation theorem does not hold.
 \end{description}
\ \\
\subsubsection*{\rm \it CASE C : NUMERICAL ANALYSIS OF PATHWAYS WITH REVERSIBLE MICHAELIS-MENTEN KINETICS}
 \begin{description}
 \item[Assumption 1.] Each enzyme in the pathway behaves
 according to reversible Michaelis-Menten enzyme kinetics. Isolated
 Michaelis-Menten enzymes do allow for enzyme saturation. Note that such kinetics
 are
 manifest in empirical studies of isolated enzymes in vitro.
 \item[Assumption 2.] When Michaelis-Menten enzymes are placed in
 the context of a multi-enzyme pathway, their kinetic
 mechanisms do not change. This is a pure assumption.
 \item[Assumption 3.] Maximal rate of input into the metabolic pathway is limited
 by a diffusion process. As in assumption 1, this is is an assumption that is empirically
 justified \cite{Tralau2000}.
\item[Numerical Result 1.] In regimes where both enzymes are
saturated the summation theorem does not hold.
\item[Numerical Result 2.]In regimes where flux is near the diffusion limited rate, the summation theorem does not hold.
 \item[Assumption 4.] Mutations can cause heritable variation in
 the catalytic turnover rate \0 k_{cat} \0 of an enzyme.
\item[Numerical Result 3.] The control of a pathway can be
switched from one enzyme to another via mutations affecting \0
k_{cat} \0. Each enzyme can be the sole controlling enzyme of a
pathway.
 \end{description}

   The main reason for the restricted applicability of the summation
   theorem is the fact that for its construction the non-linear interaction effects on flux were
   ignored.  It is precisely
   the interaction effects between enzymes that lead to the manifestation of interaction effects (~epistasis~).
    Furthermore it is the interposition of the non-linear properties of enzymes that
   allow for the evolution of control.

\section*{2.~ Mathematical Introduction}
\subsection*{2.1.~The conception of control in MCA}

   Upon devising MCA, as Kacser and Burns put it, it was necessary to
   place the understanding of individual enzyme behavior into the
   context of a ``metabolic society" composed of several enzymes \cite{Kacser1979}. In
   accordance to this goal,  one of the objectives of metabolic
   control analysis was to ascertain what role each enzyme could play
   in affecting steady state flux in a particular metabolic pathway.
   Specifically, how important each enzyme is in its ability to
   affect steady state flux via the regulation of the enzyme concentration.
   Hence the use of the control coefficient as
   \0
    C_{i}^{J}=
           { \n \frac{ \delta J}{J} / \n \frac{ \delta E_{i}}{E_{i}}
           }
\0, which is a
   non-dimensionalized sensitivity measure that shows the ratio of
   proportional change in flux to proportional change in enzyme
   concentration.

   \subsection*{2.2.~The flux summation theorem}

   One of the central tenets of metabolic control theory has been
   that the sum of the control coefficients in a pathway with $n$
   enzymes equals one. That is,
   \begin{equation}
   \nsum \cij =1. \la{st} \end{equation} Equation (\ref{st}) is
   commonly referred to as the flux summation theorem and was derived
   by Kacser and Burns using the following form of reasoning:

   Given an unbranched chain of \0n\0 enzyme catalyzed reactions, if
   we were to increase the concentration $E_{i}$ of enzyme \0i\0 by the
   proportional ratio ${\delta E_{i}}/{E_{i}}$, the change in flux due to enzyme $i$ would
   be given by rearranging (\ref{cij}) to give:
   \begin{equation}
\mbox{For all enzymes \0 i \0 :}
    ~\left( {\frac{\delta J}{J}} \right)_{i}= \left( \frac{\delta E_{i}}{E_{i}} \right) \cij  .
   \la{7b} \end{equation} If  \0\delta E_{i}/E_{i}=\alpha \0 , where
   \0\alpha\0 is a constant, one could consider the case where all
   enzyme concentrations are changed by the constant proportion
   \0\alpha\0. In such a case Kacser and Burns assume the following
   equation:
   \begin{equation}
   \frac{\de J}{J} = \nsum \left(\frac{\de J}{J}\right)_{i} \la{18}.
   \end{equation} Hence, using (\ref{18}), (\ref{7b}) and \0
   \alpha={\delta E_{i}}/ E_{i} \0 ;
   \begin{equation}
   \frac{\de J}{J} = \nsum \alpha \, \cij = \alpha \nsum \cij
   \la{19}. \end{equation}
    Separately, the assumption is made that if all enzyme
   concentrations are uniformly changed by \0 \alpha\0 , then
   \begin{equation}
   \frac{\de J}{J} =\alpha . \la{20} \end{equation}
    Substituting
   equation (\ref{20}) into (\ref{19}) we obtain equation (\ref{st}):
   \[
   \nsum \cij =1 .
   \]
For a clearer view, the Kacser and Burns derivation of (\ref{st})
can be summarized in the following conditionals:

\be
\begin{array}{l}
\left( \begin{array}{c} (\ref{7b});~ (\mbox{ Definition })
~~~~\left[~\left( {\frac{\delta J}{J}} \right)_{i}= \left(
\frac{\delta E_{i}}{E_{i}} \right)  \cij ~\right] ~\andd \\ \ \\
(\ref{18});~ (\mbox{ Assumption }) ~~~~\left[~\frac{\de J}{J} =
\nsum \left(\frac{\de J}{J}\right)_{i} ~\right] ~\andd \\ \ \\
\mbox{( Restriction )}~~~~(\forall i):
 [~{\delta E_{i}}/ E_{i}=\alpha ~]
\end{array} \right)\\ \ \\
~~~~~~~~~~~~~~~~~~~~~~~~~~~~~~~~~~~~~~~~ \Rightarrow \left(
\begin{array}{r} (\ref{19}); ~~~~\left[~ \frac{\de J}{J} = \alpha
\nsum \cij ~\right]
\end{array} \right).
\end{array}
\la{der1} \end{equation} \ \\
\be
\bea \left( \begin{array}{c} (\ref{19}); ~~~~\left[~ \frac{\de
J}{J} = \alpha \nsum \cij ~\right] ~\andd \\ \ \\ \mbox{(
Restriction )}~~~~(\forall i): [~{\delta E_{i}}/ E_{i}=\alpha ~]
\\ \ \\ (\ref{20});~ (\mbox{ Assumption })
~~~~\left[~\left(~(\forall i):
 [~{\delta E_{i}}/ E_{i}=\alpha ~]~\right) \Rightarrow \left[~\frac{\de J}{J} =\alpha ~\right]~\right]
\end{array} \right)\\ \ \\
~~~~~~~~~~~~~~~~~~~~~~~~~~~~~~~~~~~~~~~~~~~\Rightarrow \left(
\begin{array}{r} (\ref{st}); ~~~~\left[~ \nsum \cij =1 ~\right]
\end{array} \right).
\end{array}
\la{der3} \end{equation} \ \\ Note that technically even if the
assumptions \equ{18} and \equ{20} were true, \equ{st} would be
only proven for the restricted case where for every enzyme \0 i \0
: \0[~{\delta E_{i}}/ E_{i}=\alpha ~]\0.

    \section*{3.~Construction of two-enzyme pathway models}
    \subsection*{3.1.~Two-enzyme pathways as dynamical systems}
     A general dynamical model of multi-enzyme systems can be formulated  based on the
   classical model of single-substrate enzyme catalysis. Under such a scheme the
   existence of an intermediate enzyme-substrate complex is
   posited
   and a set of forward and backward reaction rates is attributed to
   each transformation. Consider a pathway in which an outside
   substrate $ u $ diffuses into the initial
   reaction compartment,where the substrate is fed into a two
   reaction pathway that comprises two successive enzyme catalyzed
   reactions. An irreversible sink step is added to the end of the
   reaction sequence in order to remove  the product.

  \[
   \mbox{\em input(u)}~\stackrel{\mbox{\tiny
   Diffusion}}{\leftrightarrow }~s_1
   ~\stackrel{\mbox{\tiny Enzyme 1}}{\leftrightarrow  }~s_2~
   \stackrel{ \mbox{\tiny Enzyme 2}}{\leftrightarrow
   }~s_3~\stackrel{\mbox{\tiny Sink}}{ \longrightarrow
   }~\mbox{\em output(o)}
   \]
   The above scenario is represented by the following kinetic model shown schematically:

\begin{equation}
\left( \begin{array}{c}
  u~\begin{matrix}  \stackrel{D}{\textstyle \longrightarrow} \\
    \stackrel{D}{\longleftarrow} \end{matrix} ~s_1 \\ \\
 e_1+ s_1~ \begin{matrix}  \stackrel{ k1 }{\textstyle \longrightarrow} \\
    \stackrel{ k2 }{\longleftarrow} \end{matrix} ~(\underline{es}_{1})
   ~\begin{matrix}  \stackrel{ k3 }{\textstyle \longrightarrow} \\
    \stackrel{ k4}{\longleftarrow} \end{matrix} ~s_2+e_1 \\ \\
  e_2+s_2 ~\begin{matrix}  \stackrel{  k5}{\textstyle
\longrightarrow} \\
    \stackrel{ k6}{\longleftarrow} \end{matrix} ~ (\underline{es}_{2})
   ~\begin{matrix}  \stackrel{ k7}{\textstyle \longrightarrow} \\
    \stackrel{ k8}{\longleftarrow} \end{matrix} ~ s_3+e_2  \\ \\
 s_3 \stackrel{ Q}{\longrightarrow} o  \\
 \end{array} \right) \la{kinetic model} \end{equation}
\

\nin where $(\underline{es}_{1})$ and $(\underline{es}_{2})$ are
the enzyme-substrate complexes, while
   $e1$ and $e2$ are the first and second enzymes catalyzing the two
   respective reactions. The input and output correspond to \0 u \0 and  \0 o \0
   respectively.
    The kinetic
   model (\ref{kinetic model})
is a class of model that can be represented as a system of
autonomous differential
   equations with input $u(t)$ and  initial
   conditions ${\bf s}(0)$ and ${{\bf e}}(0)$ such that;

    \begin{equation}
   \begin{array}{c}
      {  \dot{ {\bf s}}(t)=f_{1} \mbox{\large [}~ {{\bf s}}(t),{ {\bf e}}(t), u(t)  ~\mbox{\large ]} } \\
      {  \dot{ {\bf e}}(t)=f_{2}\mbox{\large [}~ { {\bf s}}(t),{{\bf e}}(t) ~\mbox{\large ]}  } \\
  {  \dot{o}(t)=f_{3}\mbox{\large [}~ s_{3}(t) ~\mbox{\large ]}  \label{eq1}  }
   \end{array}
   \end{equation}
   where at time $t$, the vector $ {\bf s}(t) =
   \langle ~  s_{1}(t),s_{2}(t) , s_{3}(t) ~\rangle$ is the
   state vector of substrate concentrations,   the vector
       ${\bf e}(t) =\langle  ~e_{1}(t),\underline{es}_{1}(t), e_{2}(t),\underline{es}_{2}(t) ~\rangle $
       is the state vector of enzyme and enzyme-substrate complex
       concentrations, the variable \0o(t)\0
       is the concentration associated with the output variable
        and \0 \dot{o}(t) \0 is the output flux.
                The  notation $\dot{x}$ associated with any variable $x$
       denotes the time derivative
       $dx/dt$.
     If the system reaches a steady state flux the
        output flux \0 \dot{o}(t) \0 is equivalent to the steady state flux \0 J \0 used in the MCA literature.

   \subsection*{3.2.~Characterization of two-enzyme pathways into distinct classes}
We will make the rest of our argument using
a set theroetic framework. For any given set, if we can
define the general properties of its members, then any object
which is deemed to be a member of that set will also have those
properties. A system of equations is a logical proposition. As
such we can cast our dynamical system model in a set theoretic
guise. A pathway \0 p \0 is an object. Each object has a series of
properties \0K\0 associated with it, such that \0K\lfloor p
\rfloor=\mbox{True}\0. The properties \0K\0 delineate the context
and system of equations that can describe a pathway ( For
alternative representations see Fontana \& Buss 1996 \nocite{Fontana1996}).
 Each pathway
\0 p \0 can then be treated as an element of a set with
distinct properites.
\newcounter{defs}
\setcounter{defs}{1}


\subsubsection*{Definition \arabic{defs} \addtocounter{defs}{1}:
~~\0\spoo\0 \rm \it \0\equiv\0 Set of two enzyme pathways with one
input and one output. }

 There are no strict stipulations on the
stoichiometry nor on the specifics of
 enzyme kinetic mechanisms for membership in \0 \spoo \0 ( see Appendix A ).


\ \\
\subsubsection*{Definition \arabic{defs}:~~\0\spuJ\equiv\0
 \rm \it Set of two-enzyme pathways with constant input and steady-state
 flux}
 The set
$\spuJ $ is defined as the subset \0 \spuJ \subset \spoo \0 such
that \0 p \in \spuJ \0 if and only if the following four
conditions hold:
\begin{enumerate}
\item
 \0 p \in \spoo \0
\item
Irrespective of the enzyme kinetic mechanisms, \0 p \0 has the
stoichiometry of a two-enzyme sequential pathway as exhibited in
model (\ref{kinetic model}).
\item
When driven by a constant input \0 u \0, \0p\0 exhibits a unique
steady state output flux \0 J^*\0 that can be associated with each
pair of total enzyme concentrations \0 \langle E_1,E_2 \rangle \0.
\item
For any error margin \0 \err > 0 \0, the steady state output flux
\0J^*\0 can be
   approximated by using the total enzyme
concentrations \0 \langle E_1,E_2 \rangle \0 and constant input
\0u\0 as arguments for a computable function \0 g_\varepsilon \0
such that \0 J=g_\varepsilon [u,E_1,E_2] \0 and \0 | J-J^*
|<\varepsilon \0 .
\end{enumerate}

 Items 2 to 4 are
elaborated more precisely in Definitions \arabic{defs}.2 to
\arabic{defs}.4 in Appendix A. Note that the MCA approach holds
informal versions of these propositions as premises, without which
talking about the steady-state flux \0J\0 and the summation
theorem would not be possible.

\ \\ \addtocounter{defs}{1}

\subsubsection*{Definition \arabic{defs}:~~\0\spuDJ\equiv\0 \rm
\it Set of diffusion limited two-enzyme pathways with steady-state
flux}
 The set $\spuDJ $ is defined as the subset \0 \spuDJ
\subset \spuJ \0 such that \0 p \in \spuDJ \0 if and only if the
following two conditions hold:
\begin{enumerate}
\item \0 p \in \spuJ \0
\item The net input flux from \0 u(t) \0 into the system is governed
by a diffusion-limited process such that \0 Netstartflux=
D\left(\,u(t)-s_{1}(t)\,\right) \0 where \0D\0 is a diffusion constant.
\end{enumerate}

Item 2 is more precisely defined in definition \arabic{defs}.2 in
Appendix A.

\ \\ \addtocounter{defs}{1}

\subsubsection*{Definition \arabic{defs}:~~\0\spuVJ\equiv\0 \rm
\it Set of  two-enzyme pathways with steady-state flux and
kinetics that allows for enzyme saturation.}
 The set $\spuVJ $ is defined as the subset \0 \spuVJ
\subset \spuJ \0 such that \0 p \in \spuVJ \0 if and only if the
following two conditions hold:
\begin{enumerate}
\item \0 p \in \spuJ \0
\item For every enzyme \0i\0, there exists a constant \0 k_{cat(i)}\geq 0 \0  such that for all values of \0 E_i\geq0
\0 the steady state flux is consistent with the relation \0 J\leq
E_i\,k_{cat(i)} \0.
\end{enumerate}

Item 2 is more precisely defined in definition \arabic{defs}.2 in
Appendix A.

\ \\ \addtocounter{defs}{1}

\subsubsection*{Definition \arabic{defs}:~~\0\spuDVJ\equiv\0 \rm
\it Set of  diffusion limited two-enzyme pathways with
steady-state flux and kinetics that allows for enzyme saturation.}
 The set $\spuDVJ $ is defined as the intersection \[
\spuDVJ=(\spuDJ  \cap  \spuVJ) \subset \spuJ . \]
\ \\

  As an intuitive guide, the order of the generality
of the representations discussed so far, in order of higher
generality are:
 \0  \spuDVJ \subset \spuDJ \subset \spuJ \subset \spoo  \0 and
\0  \spuDVJ \subset \spuVJ \subset \spuJ \subset \spoo  \0 (See
Figure \ref{fig2.0}).

\section*{4.~ Problems with the summation theroem}
The definitions from the previous section have a series of logical
implications. Here we present these implications and explain their
significance. The proofs for these propositions are given in
Appendix B.
\subsection*{4.1.~Pathways with a constant input and a steady-state flux \0( \spuJ )\0}


\subsubsection*{Proposition {\ref{theorem1}}}
\em For any pathway of class $\spuJ$, the flux summation theorem
implies that for any given ratio of enzymes ~\0 0\leq (E_1/E_2)
\leq\infty \0~, there exists a \0 C_\theta \0 such that flux is a
linear function of the magnitude of the vector \0 \langle E_1,E_2
\rangle \0.
\be
\bea (\forall p\in\spuJ ):(\forall \theta ~|~ \tan\theta =
E_1/E_2): (\exists C_\theta):\\ \left[~~ \ctwo =1
~~\Rightarrow~~ J=\|\langle E_1,E_2 \rangle\|\,C_\theta
~~\right]. \ena \la{theorem1}
\end{equation}
\ \\ \rm The meaning of proposition \ref{theorem1} is very simple.
If a pathway were to be of class \0 \spuJ \0 and were to obey the
flux summation theorem the following would hold: the flux would
behave in a linear fashion with respect to tandem changes in which
both enzyme concentrations are changed by the same percentage. It
is a short few steps (Appendix B) to show that
\be
\forall \nakedalphamove:\,\,
 \left[\frac{\de J}{J}
=\alpha ~\ifif~ J=\|\langle E_1,E_2 \rangle\|\,C_\theta
\right].\la{linequiv}
\end{equation}
Proposition \equ{linequiv} is equivalent to a statement of
homogeneity (~See Giersch 1988~), such that if \0  J=g[E_1,E_2] \0
then
\be
g[~(1+\alpha)E_1,\, (1+\alpha)E_2~] = (1+\alpha)g[~E_1,\,E_2~].
\end{equation}

 For example
consider two enzymes whose genes are located on the same operon.
In addition consider the case in which such a cell is in a state
in which the concentrations of the two enzymes are initially
negligible. When the operon is turned on, each enzyme will have a
particular rate of expression. If the ratio between those rates of
expression remains constant, then the flux summation theorem would
imply that the flux would increase linearly without reaching a
saturation plateau. This is what \0 J=\|\langle E_1,E_2
\rangle\|\,C_\theta \0 implies, where \0 \| \langle E_1,E_2\rangle
\|= \sqrt{E_{1}^2+E_{2}^2}\0.  Hence, in a pathway of class \0
\spuJ \0, in which we have made no assertion about the particulars
of the enzyme-kinetic equations or parameters, the flux summation
theorem assumes a particular type of dynamic behavior. A flux that increases linearly without saturation is a very special case and such a condition can
only hold
in a restricted subset of \0 \spuJ \0.
  In fact, as proposition {\ref{theorem2}} will attest,
  even for a simple
 pathway with the sole kinetic imposition of a diffusion barrier,
 the flux summation theorem fails.

\subsection*{4.2.~Pathways with a diffusion barrier \0( \spuDJ )\0}
\subsubsection*{Proposition {\ref{theorem2}}}
\em For any pathway of class $\spuDJ$, the flux summation theorem
is false.\rm
\begin{equation}
   (\every p \in \spuDJ):
   \left[~~(\forall \de E_1,\,\de E_2 ):~(\exists E_1,E_2 \in \real{1}_{\geq 0}):~
  \left[~ \ctwo \neq 1
~\right]  ~~\right]
   \la{theorem2}\end{equation}
\ \\ Proposition \ref{theorem2} basically states
 the general result that any pathway
which has a diffusion barrier at a point before its first step
will not obey the summation theorem. This includes almost any
metabolic pathway that takes place in the confines of a cell. In
fact it  can be shown that any pathway that exhibits a plateauing
effect on its flux surface will not obey the summation theorem.
The proof of this theorem is intuitively simple (Appendix B). It
is basically dependent on the conditional relation
\be
 \ctwo =1 ~~~\Rightarrow~~~ J=\|\langle E_1,E_2
\rangle\|\,C_\theta \la{bicond}\end{equation}
 from \equ{theorem1}. For any pathway class
that is a subset of \0 \spuJ \0, if the pathways  do not obey the
right hand side of \ref{bicond} then they do not obey the left
hand side either. For \0 \spuDJ \0 one can show that \0
J=\|\langle E_1,E_2 \rangle\|\,C_\theta \0 does not hold when \0
C_{\theta} > 0 \0. This is because the diffusion barrier will
impede flux from increasing indefinitley. This implies that \0
\nsum \cij =1  \0 cannot be true for all enzyme concentrations.
  This falsehood applies to all
 diffusion-limited flux surfaces except for a
 disfunctional cell with flux surface \0 J=0 \0.
Proposition \ref{theorem2} is due to the fact that control is not only dependent on the enzymatic steps in a pathway, but also the diffusion steps.

\subsection*{4.3.~Pathways that allow for enzyme saturation \0( \spuVJ )\0}
\subsubsection*{Proposition {\ref{theorem4}}}
\em For any pathway of class $\spuVJ$, if both enzymes approach
saturation and changes in enzyme concentration are positive, then the sum of control coefficients  approaches zero.
\rm
\be
(\every p \in \spuVJ):
   \left[~(\forall\, \alldelta):~
   \limit{Sat_1,Sat_2}{1} \left( \ctwo \right)=0 ~~\right].
\la{theorem4}\end{equation} \ \\ Whenever two enzymes are near
saturation, a situation is created whereupon the flux effects of
increasing the concentration for one enzyme is restrained by the
other enzyme. Hence whenever two enzymes approach saturation
simultaneously, the sum of their control coefficients approaches
zero and the flux summation theorem does not hold ( see appendix B
). In fact it is not too difficult to extend the proof  such that
for a sequential pathway of any length, the sum of all control
coefficients will approach zero whenever any two enzymes approach
saturation.

\subsubsection*{Proposition {\ref{theorem4_2}}}
\em For any pathway of class $\spuVJ$ in which a decrease in  enzyme concentration increases saturation, if both enzymes approach
saturation and changes in enzyme concentration are negative,  then the sum of control coefficients  approaches two.
\rm
\nin  \ \\ \ \\
For all intervals \0  a_i < E_i < b_i \0:   \\
If ~\0 p \in \spuVJ \0~ and \0 ~\forall i \0 : (~\0  \frac{ \partial Sat_i} {E_i} \geq 0 \0 ~~and~~
\0 a_i \geq 0 \0  ~~and~~
\0 \de E_i<0 \0 ~)~
~~then:
\be
 \limit{Sat_1,Sat_2}{1}{\left( C_{1}^{J} + C_{2}^{J} \right) = 2 }.
\la{theorem4_2}\end{equation}
\ \\ In such situations the pathway is no longer robust to decreases in enzyme concentration. Hence dominance is not an inherent property of these pathways. It is simple to extend this result to sequential pathways of any length, in which the sum of control coefficients will approach the the number of enzymes which are approaching saturation.

From propositions  {\ref{theorem2}},  {\ref{theorem4}} and {\ref{theorem4}} it also
follows that the summation theorem is also false in pathways of class
\0 \spuDVJ \0.

\subsection*{4.5.~Enzyme interactions in \0 \spuJ \0}
The motivation for analyzing multi-enzyme pathways is that they
behave differently from                                                                                                                                                                          single-enzymes.
 The natural question is to ask what are the direct or indirect
interactions between enzymes that lead to the difference between
single-enzyme and multi-enzyme pathway behavior. Interaction does
not necessarily have to signify
 direct physical interaction between two enzymes. It can occur via
 the effect of one enzyme on the concentration of pathway intermediates that in turn
 physically interact with another enzyme. In equation (6)
which was one of the assumptions that Kacser and Burns made in
deriving the summation theorem, the proposition \0  \frac{\de
J}{J} = \nsum \left(\frac{\de J}{J}\right)_{i} \0 amounts to an
assumption that enzymes do not exhibit interactions. Here
interaction is specifically referring to the  mutual effect that
the two enzymes would have on each-other's ability to affect
pathway flux:

\subsubsection*{Proposition {\ref{theorem3}}}
\em For any pathway of class $\spuJ$, the proposition
   \0 \left[~\frac{\de J}{J} = \nsum \left(\frac{\de
J}{J}\right)_{i} ~\right] \0 (\ref{18}) and the restriction \0
\alphamove \0 imply that
 the change in flux due to
   change of total enzyme concentration at one locus is independent of total enzyme
   concentration at the second locus in the pathway.\nopagebreak
   \begin{equation}
   \bea
   (\every p \in \spuJ):\quantone: \\
   \left[~~\left[~\forall \alphamove:~~
   \frac{\de J}{J} = \nsumtwo \left(\frac{\de
J}{J}\right)_{i}\right] ~~\Rightarrow~~ \dpj=0 ~~~\right].
   \ena
   \la{theorem3}
   \end{equation}
\ \\ \rm  The assumption of independence of effects is something
that surreptitiously goes against the motivation for developing
MCA. Nonetheless the
assumption of no enzyme interactions is built into the summation
theorem from the beginning. In fact:

\subsubsection*{Proposition {\ref{theorem5}}}
\em For any pathway of class $\spuJ$, the summation theorem
implies that
 the change in flux due to
   change of total enzyme concentration at one locus is independent of total enzyme
   concentration at the second locus in the pathway.\nopagebreak
   \begin{equation}
   \bea
   (\every p \in \spuJ):\quantone:  \\
     ~~~~~~~~~\left[~\left[(\forall \de E_1,\,\de E_2 ) :~\ctwo = 1~\right]  ~~\Rightarrow~~ \dpj=0 ~~~\right].
   \ena
   \la{theorem5}
   \end{equation}
\ \\ \rm This means that the summation theorem implies the
inexistence of epistasis. It immediately follows that
\begin{equation}
\bea
   (\every p \in \spuDJ): ~\forall \left(E_1,E_2 \in \real{1}_{\geq 0}~|~\dpj\neq 0~\right):~\\
   ~~~~\left[~~(\exists \de E_1,\,\de E_2):~
  \left[~ \ctwo \neq 1
~\right]  ~~\right].
   \la{theorem5two}
\ena
   \end{equation}
Hence in any regime where epistasis is present the summation
theorem is violated. Furthermore:
\subsubsection*{Proposition {\ref{theorem6}}}
\em For any pathway of class $\spuJ$, the summation theorem holds
if and only if flux is a linear function of enzyme
concentrations.\nopagebreak
   \begin{equation}
   \bea
   (\every p \in \spuJ):\quantone:  \\
     ~~~~~~~~~\left[~\left[(\forall \de E_1,\,\de E_2) :~\ctwo = 1~\right]  ~~\ifif~~
     J = E_1\,c_1 + E_2\,c_2 ~~~\right].
   \ena
   \la{theorem6}
   \end{equation}
\ \\ \rm

 In short, the summation theorem imposes
 conditions on enzyme interactions that are neither explicitly argued for nor conceptually
 justified.

  \section*{5.~ Numerical analysis of a pathway of class \0 \spuDVJ\0}

Here we use numerical simulations for a particular case in order to verify
the analytical results from the previous section.
Consider a case where the enzymes exhibit reversible
Michaelis-Menten kinetics. A diffusion barrier is included at the
beginning of the pathway.
 During the interval of
   analysis $0\leq t\leq t_f$, the input remains constant such that
   $\dot {\bar {u}}(t)=0 $ (e.g. approximation of large nutrient pool and/or steady state conditions
   of a chemostat). In such a case the full form of the system of
   differential equations ($\ref{eq1}$)
 combined with the kinetic model (\ref{kinetic model}) are:

   \begin{equation}
       \left(
       \begin{array}{l}
       \frac {\n d[s_{1}]}{\n dt}=
           - D\,{[s_1]}   - {k1}\,{[e_1]}\,{[s_1]} + D\,[u] + {k2}\,{[\underline{es}_{1}]} \\\\
       \frac {\n d{[\underline{es}_{1}]}}{\n dt}=
           {k1}\,{e_1}\,{[s_1]} - {k2}\,{[\underline{es}_{1}]} - {k3}\,{[\underline{es}_{1}]} + {k4}\,{[e_1]}\,{[s_2]} \\\\
       \frac {\n d[s_{2}]}{\n dt}=
           {k3}\,{[\underline{es}_{1}]} - {k4}\,{[e_1]}\,{[s_2]} - {k5}\,{[e_2]}\,{[s_2]} + {k6}\,{[\underline{es}_{2}]} \\\\
       \frac {\n d[\underline{es}_{2}]}{\n dt}=
           {k5}\,{[e_2]}\,{[s_2]} - {k6}\,{[\underline{es}_{2}]} - {k7}\,{[\underline{es}_{2}]} + {k8}\,{[e_2]}\,{[s_3]} \\\\
       \frac {\n d[s_{3}]}{\n dt}=
           {k7}\,{[\underline{es}_{2}]} - Q\,{[s_3]} - {k8}\,{[e_2]}\,{[s_3]} \\\\
       \frac {\n d[e_{1}]}{\n dt}=
           - {k1}\,{[e_1]}\,{[s_1]} + {k2}\,{[\underline{es}_{1}]} + {k3}\,{[\underline{es}_{1}]} - {k4}\,{[e_1]}\,{[s_2]} \\\\
       \frac {\n d[e_{2}]}{\n dt}=
           - {k5}\,{[e_2]}\,{[s_2]} + {k6}\,{[\underline{es}_{2}]} + {k7}\,{[\underline{es}_{2}]} - {k8}\,{[e_2]}\,{[s_3]} \\\\
       \frac {\n d[o]}{\n dt}=Q\,{[s_3]} \\\\
       \end{array} \right) \label{odes}
   \end{equation}
   where the postfix \0(t)\0 for each state variable is omitted.

The system of differential equations \equ {odes} will serve as the
basis for the examination of a particular case. For the present
analysis we are interested in the steady states of the dynamical
system. At steady state, as defined in appendix A, the derivatives
\0 \dot{ {\bf s}}(t) \mbox{ and } \dot{ {\bf e}}(t) \0 are equal
to zero. Consider a case where the equations in \equ{odes} serve
as the basis for the functions \0 f_1 \0, \0 f_2 \0 and \0 f_3 \0
with the stoichiometric constraint that
\begin{equation}
\begin{array}{c}
{[E_{1}]=[e_{1}]+[\underline{es}_{1}] } \\
{[E_{2}]=[e_{2}]+[\underline{es}_{2}]}.
\end{array}
\end{equation}
In such a case, the steady state solution can be obtained in an
implicit form ( See appendix C ). The implicit equations can then
be solved numerically by Newton-Raphson based
 algorithms that converge to the flux surface shown in Figure \ref{fig2.1}.
On a cursory level the surface shows the plateau effect familiar
in most models of
 two-enzyme pathways. The plateau arises due to the
limiting effects of the diffusion barrier.
 We will refer to the pathway based on
 equations \equ{odes} as pathway \0 p_{\ref{odes}} \0. It is not difficult to discern
 that pathway \0 p_{\ref{odes}} \0 is a member of the set \0
 \spuDVJ \0. It satisfies the  conditions required in definitions 1 to 5.

\subsection*{5.1.~Test of the summation theorem}
Given the flux surface on \0 p_{\ref{odes}}\in \spuDVJ \0, the
flux summation theorem can be tested directly. Figures
\ref{fig2.2} and \ref{fig2.2.2} show surfaces for for discrete
versions of the sum \0 \nsum \cij \0 for \0 p_{\ref{odes}} \0. It
is evident from these examples that in the pathway \0
p_{\ref{odes}} \0 there is no evidence of an invariant relation
 in which the control coefficients sum to the predetermined
 value of one. Figures \ref{fig2.2} and
\ref{fig2.2.2} serve as illustrations of propositions
{\ref{theorem2}}, {\ref{theorem4}} and {\ref{theorem4_2}}. The
summation theorem fails in the regimes where flux is
 near its diffusion-limited maximal rate or in areas where both enzymes
  are near saturation. The regions where both enzymes approach saturation are
 shown in Figure \ref{fig2.3}.

\subsection*{5.2.~Epistatic interactions between enzymes}
The question of  the extent to which enzymes have interactive
effects on flux can also be addressed directly using the pathway
\0 p_{\ref{odes}} \0 as an example. In Figure \ref{fig2.2-2}
 the mixed partial difference \0
{\delta^{2} {J}}/{\delta{E_1}\delta{E_2}} \0 is  used as a measure
of epistatic interactions between enzymes (proposition
\ref{theorem3}). Note that the region where the epistatic
interactions increase coincides with a region where the summation
theorem fails and  both enzymes are approaching saturation.

\subsection*{5.3.~Rate limiting enzymes and the switching of control between enzymes}
One of the claims from MCA has been that there is no such thing as
a rate limiting enzyme in a pathway. The associated claim is that
control is mostly shared between enzymes in a pathway \cite{Kacser1995b}.
 The results of our numerical simulations of \0
p_{\ref{odes}} \0 do not support this claim. Figure \ref{fig2.4}
shows the numerical results for two pathways with different \0
k_{cat} \0 configurations. In the range of concentration regimes
\0 \langle E_1,E_2 \rangle \0 examined, most regimes manifest a
sole enzyme as the controlling enzyme for a given pathway. In
regimes where flux is near the diffusion limited maximal flux, no
enzyme has control. The only region where the enzymes share
control is where the summation theorem fails. In fact in the
regions that the summation theorem holds, the sum
\0\sum_{i=1}^{2}\cij=1\0 is basically the value \0 C^{J}_{z}\0 for
the sole controlling enzyme \0 z \0 at that regime. Furthermore
the identity of the controlling enzyme at a given regime is not
invariant. The identity of the controlling enzyme can change by
changing the \0 k_{cat} \0 values of the enzymes involved (e.g. by
mutation ). In Figure \ref{fig2.4} the switch in control is
clearly apparent by comparing the values of \0 C_{2}^{J}\0 between
the two \0 k_{cat} \0 configurations.

  \section*{6.~Discussion and conclusion}

The main intent of the present paper has been to present an
analysis and critique of some of the main biological conclusions
derived from MCA. Our results strongly support initial objections
that were made with regards to the MCA perspective on the
biochemical nature of dominance ( Cornish-Bowden 1987, Savageau
1992. See also Mayo \& Burger 1997 \nocite{Mayo1997}).
 At the core of the
problem of multienzyme systems is interaction effects between
enzymes. In order to address these interactions, the non-linear
properties of enzyme catalysis cannot be ignored. These non-linear
interactions play a major role in the phenotypic characteristics
of metabolic physiology. Our analysis shows that for discrete
changes of any magnitude, the flux summation theorem does not
hold. This means that there are no a-priori constraints that would
require the magnitude of mutational effects to be in a low range.
This leaves open the possibility for two phenomena which had been
previously rejected in MCA. In the first place, rate limiting
steps or controlling enzymes do not have to be rare, and their
identity can change by evolution. Secondly, phenotypic robustness
with respect to mutations is not an automatic property of
metabolic pathways; the degree of robustness can be modified
through the evolution of kinetic properties. These are
implications that we will examine in a later work.

\pagebreak
\section*{Appendices}
\setcounter{defs}{1}
\section*{A.~~Definitions}

In this section some of the conditions proposed in definitions 1,2,
3 and 4 are formulated in a more precise language as definitions
2.2, 2.3, 2.4, 3.2 and 4.2. In these definitions quantifiers are
encapsulated in parenthesis and propositions in square brackets.
The symbol ``\0 \wedge \0" denotes the boolean expression ``and".
 The  symbol ``\0
\real{1}_{\geq 0} \0" denotes the real numbers greater than or
equal to zero. Ordered tuples and vectors are encapsulated by the
symbols ``\0 \langle ~\rangle\0". The property \0 K \0 associated
with an object \0 p \0  is denoted as \0 (\mbox{``name of
property"})\lfloor p \rfloor \0.

\subsection*{Definition \arabic{defs}}

\begin{equation}
 \begin{array}{l} \ \\
( \forall p \in \spoo):
\begin{array}{l}
 ~ p \Rightarrow
  \left[
\begin{array}{l}
{\dot{ {\bf s}}(t)=f_{1} [
 {{\bf s}}(t),{ {\bf e}}(t), u(t) ] } ~~\wedge \\
 {\dot{ {\bf e}}(t)=f_{2}[{ {\bf s}}(t),{{\bf e}}(t) ]  } ~~\wedge \\
 {\dot{o}(t)=f_{3}[ s_{3}(t)] } \andd\\
\left( u(t),o(t) \in \real{1} _{\geq 0} \right) \andd
\left( {\bf e}(0)\in \real{4} _{\geq 0} \right)
\andd \left( {\bf s}(0) \in \real{3} _{\geq 0} \right)
\end{array} \right] \\
\end{array}
\end{array}
\la{def2} \end{equation}

\addtocounter{defs}{1}

\subsection*{Definition \arabic{defs}.2}

The proposition \0\A \0 imposes both organizational and kinetic
constraints on the pathway \0 p\in \spuJ\0 by designating the
stoichiometric properties of the pathway. Such a proposition is
necessary for an abstract representation of physical mass balance.
\begin{equation}
\begin{array}{l}
\A \equiv \\ ~~\begin{array}{l} (\,\forall (t,u(t),o(t))\in
\real{1} _{\geq 0}\,): (\forall {\bf e}(0)\in \real{4} _{\geq 0} )
: ( \forall {\bf s}(0) \in \real{3} _{\geq 0}): \\ ~\left(
\begin{array}{l}
 \left[
\begin{array}{l}
  \begin{array}{l}
   {[~\dot{E}_{1}(t)=\dot{E}_{2}(t)=0~]}~~~~\wedge \\
   {[~E_1=E_{1}(t)=e_{1}(t)+\underline{es}_{1}(t)~]} ~~\wedge \\
   {[~E_2=E_{2}(t)=e_{2}(t)+\underline{es}_{2}(t)~]} ~~\wedge \\
   {[~E_1\geq0 ~\wedge~   E_2\geq0~] }
  \end{array}
 \end{array} \right] ~~~\andd \\ \ \\
\left(\begin{array}{l} \left[\begin{array}{l}
{\xi(t)=s_1(t)+2\underline{es}_{1}(t)+s_2(t)+2\underline{es}_{2}(t)}
\\ {~~~~~~~~~~+s_3(t)+e_1(t)+e_2(t)} \end{array} \right] {\andd}
\\ \
\\ \left( \begin{array}{l}
 (~\exists I_{in}(t),I_{rev}(t) \in \real{1}_{\geq0}~) :\\
~{[~~\dot{\xi}(t)=I_{in}(t)- I_{rev}(t) -\dot{o}(t)~~]}
\end{array} \right)
\end{array}\right)
 \end{array} \right)
\end{array}
\end{array}
\la{stoichiometry}
\end{equation}
where the pair $\langle E_{1},E_{2}\rangle
\in\mathbb{R}^{2}_{\geq0}$ are the total enzyme concentrations for
any pathway $p$  such that \0\langle E_1, E_2\rangle\0 does not
change for each particular vector of starting conditions and that
for a given set of initial conditions, the concentrations \0E_1
\mbox{ and } E_2\0 are to be determined at any time $t$ by summing
the free-enzyme and enzyme-substrate complex concentrations for
each enzyme respectively. The variable \0\xi(t)\0 needs some
explanation. \0\xi(t)\0 determines the ``stoichiometric content''
of the pathway in the sense that the coefficients \0c_i\0 in the
equation
\[
{\xi(t)=c_1s_1(t)+c_2\underline{es}_{1}(t)+c_3s_2(t)+c_4\underline{es}_{2}(t)
+c_5s_3(t)+c_6e_1(t)+c_7e_2(t)} \] are determined by whether in
reference to the input molecule, the transformations leading to
the macro-molecules associated with each coefficient are fusions,
cleavages or one-one transformations. The values of the \0 c_i\0
in (\ref{stoichiometry}) correspond to the kinetic model
(\ref{kinetic model}). The equation for \0 \dot{\xi}(t) \0  also
deserves attention.
 Each metabolic pathway can be characterized
 such that  \0 \dot{\xi}(t) \0 is determined by the additive combination
 of two values at time \0t\0:
 \[ \dot{\xi}(t) = \mbox{\em  Netstartflux(t) - Netendflux(t)} \]
 which in turn can be subdivided as
 \[
 \bea
 { \mbox{\em  Netstartflux(t) = \0 I_{in}(t) - I_{rev}(t)\0 \rm~~~~~~~~~~~~and} }
 \\ { \mbox{\em Netendflux(t) =\0 O_{out}(t) - O_{rev}(t) \0} }
 \end{array}
 \]
where \0 I_{in}(t) \0 is the flux entering the system from the
input variable \0u(t)\0, \0 I_{rev}(t) \0 is the flux exiting the
system and going back into \0u(t)\0, \0 O_{out}(t) \0
 is the flux exiting the system into the output variable \0o(t)\0 and
 \0 O_{rev}(t) \0 is the flux entering back into the system from
 \0o(t)\0. For a kinetic model with an irreversible sink step,
 \0 O_{rev}(t)=0 \0. For a system
 where the output variable \0{o}(t) \0 is only driven by a single pathway \0p\0,
 \0 O_{out}(t)=\dot{o}(t) \0. Finally for \0\spuJ\0 the only stipulations made
 for the two components of \em Netstartflux(t) \rm is that they must both be greater or equal to zero
 such that \0 I_{in}(t),I_{rev}(t) \in \real{1}_{\geq0}\0 and
 \0 \mbox{\em Netstartflux(t)}=I_{in}(t)-I_{rev}(t) \0.

\subsection*{Definition \arabic{defs}.3}

\indent For a pathway \0 p\in \spuJ\0 the logical proposition \\
 \0\B\0 states that given an input \0u(t)\0, if
\0u(t)\0 remains constant then the pathway will reach a unique
steady state flux \0J^* \in \real{1}_{\geq0}\0 that is dependent
on total enzyme concentrations \0E_1 \mbox{ and }E_2\0 such that:
\begin{equation}
\begin{array}{l}
\B \equiv \\ ~(\,\forall u(0)\in \real{1} _{\geq 0}\,):
 ( \forall E_1(0),E_2(0) \in \real{1} _{\geq0}):\\
~~(~\exists J^*\in \real{1}_{\geq0}~): (\,\forall o(t)\in \real{1}
_{\geq 0}\,):
  ( \forall s_1(0),s_2(0),s_3(0) \in \real{1} _{\geq0}):\\
~~~\left(\begin{array}{l} \left(\begin{array}{l} (\forall t \in
\real{1}_{\geq0}):\\ ~~{[~\dot{u}(t)=0~]}
\end{array}\right)
~\Rightarrow
 ~\left[   \begin{array}{l}
  \limit{t}{\infty} \dot{{\bf s}}(t)=\langle 0,0,0 \rangle  \andd \\
  \limit{t}{\infty} \dot{{\bf e}}(t)=\langle 0,0,0,0 \rangle  \andd \\
 \limit{t}{\infty} \ddot{o}(t)=0  \andd \\
  \limit{t}{\infty} \dot{o}(t)=J^*
 \end{array} \right]
\end{array} \right)
\end{array}
\la{sdef3b}\end{equation}

\nin Note that at steady state \0 \dot{\xi}(t)=0 \0.
\subsection*{Definition \arabic{defs}.4}

The proposition \0\C\0 states that an approximation
 \0J\0 of \0J^*\0  is computable as a function of
the initial input and total enzyme concentrations:
 \begin{equation}
\begin{array}{l}
{\C \equiv} \\
 \left(
\begin{array}{l}
{(~\forall \varepsilon \in \real{1}_{>0}~): (\exists
g_{\varepsilon}: \fset{3}{1}):}  \\ {(\,\forall (u(0),o(t))\in
\real{1} _{\geq 0} \,): (\forall {\bf e}(0)\in \real{4} _{\geq 0}
) :  ( \forall {\bf s}(0) \in \real{3} _{\geq 0}):}\\ \left(
\begin{array}{l}
  ~~{[~ J=g_{\varepsilon}[u(0),E_{1},E_{2}] ~\wedge
~|J-J^{*}|\leq\varepsilon ~]}
\end{array}\right) \\
\end{array} \right)
\end{array}
\la{sdef3c}
\end{equation}

\nin The above definition means that for any error margin
 \0\varepsilon\geq 0\0, there exists at-least one effectively computable
function \0 g_{\varepsilon} \0 that maps to the approximate value
\0J\0 by using the total enzyme concentrations and initial input
as arguments. Note that \0 g_{\varepsilon} \0 as an effectively
computable function refers to both numerical and analytic
functions that can produce an approximation \0J\0. The assumption
of computability is associated with the assumption that there is a
rule based mechanistic relation between \0 E_1\0, \0 E_2\0 and \0
u \0.
  There is a strong assumption inherent in
the definitions of \0 g_\varepsilon \0 in \0\spuJ\0: that for any
pathway \0 p \in \spuJ \0 a steady state flux exists and can be
determined using only \0u(0)\0, \0E_1\0 and \0E_2\0 as arguments.
 That is, except for the constraints imposed by \0 u(0),E_1\0 and \0E_2\0
 the
steady state is independent of the specifics of starting
conditions. Note that for classes of pathways that exhibit
multiple equilibria the trajectories become important and all the
starting conditions have to be included as arguments in a newly
defined computable function other than \0 g_\varepsilon \0, which
would be embedded in a system other than \0\spuJ\0.

\addtocounter{defs}{1}

\subsection*{Definition \arabic{defs}.2}
  \begin{equation}
\begin{array}{l}
\D \equiv \\
 ~~(\exists D \in \real{1}_{\geq0}):\\
 ~~~~(\,\forall (t,u(t),o(t))\in \real{1} _{\geq 0}\,):
(\forall \bf{e}(0)\in \real{4} _{\geq 0} ) : ( \forall \bf{s}(0)
\in \real{3} _{\geq 0}): \\ ~~~~~\left(\bea {[~I_{in}(t)=D \,
u(t)~] ~\andd~} \\ {[~I_{rev}(t)= D \, s_{1}(t)~]}
\end{array}\right)
\end{array}   \la{diffusion limited}
\end{equation}
where \0 D\0 is the diffusion constant.

\addtocounter{defs}{1}

\subsection*{Definition \arabic{defs}.2}
  \begin{equation}
\begin{array}{l}
\E \equiv \\
 ~~(\exists (k_{cat(1)},k_{cat(2)}) \in \real{1}_{\geq0}):
 ( \forall (E_1,E_2)\in \real{1} _{\geq 0}):\\
 ~~~~(\,\forall (t,u(t),o(t))\in \real{1} _{\geq 0}\,):
(\forall \bf{e}(0)\in \real{4} _{\geq 0} ) : ( \forall \bf{s}(0)
\in \real{3} _{\geq 0}): \\
 ~~~~~\left(
 \bea {[~J \leq E_1 k_{cat(1)}~] ~\andd~} \\
  {[~J \leq E_2 k_{cat(2)}~]}
\end{array}\right)
\end{array}   \la{saturable enzymes}
\end{equation}

\ \\
\section*{B.~~Proofs}

%
 \subsection*{B.1~~Proof of proposition {\ref{theorem1}}}
We start with the summation theorem \0 \nsum \cij =1 \0. Using the
definition of \0 \cij \0 from (\ref{cij}) we have
\be
(\forall E_1,E_2 \in \real{1}_{> 0}):~~~\disspart{J}{E_1} \,
\frac{E_1}{J}+ \disspart{J}{E_2} \, \frac{E_2}{J} =1. \la{aone}
\end{equation} The continuous version of (\ref{aone}) is \0
\limit{\de E_1,\, \de E_2}{0} \left(\disspart{J}{E_1}\, {E_1}+
\disspart{J}{E_2} \,{E_2}\right) = J \0 which is equivalent to
\be
\spart{J}{E_1} \,{E_1}+ \spart{J}{E_2} \,{E_2}  =J.
\la{atwo}\end{equation} Equation (\ref{atwo}) is a first order
quasi-linear partial differential equation with general solution
\be
J= E_2 \, f\left[ \frac{E_1}{E_2} \right]=E_1 \, f'\left[
\frac{E_2}{E_1} \right], \la{athree}\end{equation} where \0
f,f':\real{1}\mapsto\real{1} \0 (see for example Rade \&
Westergren 1995, p230). For each fixed ratio of \0 E_1/E_2 \0, the
vector \0 \langle E_1,E_2 \rangle \0 forms a fixed angle \0 \theta
\0 with the coordinate axes \0E_2\0 and \0E_1 \0 such that \0
E_1/E_2 = \tan\theta \0. Hence we can formulate the following
proposition:
\be
\begin{array}{c}
(\exists g_{st}: \fset{2}{1}): (\forall E_1,E_2 \in \real{1}_{>
0}) : (\exists f_{st}: \fset{1}{1}):\\ J~=~
g_{st}[E_1,E_2]~=~E_2\,f_{st}[\tan\theta]
~=~h\,\cos\theta\,f_{st}[\tan\theta]
\end{array}
\la{aa37} \end{equation} where \0h\0 is the norm of the vector
\0\langle E_1,E_2\rangle\0 such that \0 h=\| \langle
E_1,E_2\rangle \|= \sqrt{E_{1}^2+E_{2}^2}\, \0. Given that for
every \0 \theta \0 there exists a \0C_\theta\0 such that \0
C_\theta=\cos \theta \,\cdot f_{st}[\tan \theta] \0, we have
\be
(\forall \theta \mid 0\leq\theta\leq{\pi}/{2}):(\exists C_\theta):
[~J=h\,C_\theta~]. \la{afour}\end{equation}
\0~~~~~~~~~~~~~~~~~~~~~~~~~~~
~~~~~~~~~~~~~~~~~~~~~~~~~~~~~~~~~~~~~~~~~~~~~~~~~~~~~\blacksquare
{\bf ~proposition~ {\ref{theorem1}}}\0

\subsection*{B.2~~Proof of Proposition \ref{linequiv}}
The proof of the proposition \0 \left[ \frac{\de J}{J} =\alpha
~~\ifif~~ J=\|\langle E_1,E_2 \rangle\|\,C_\theta \right] \0 is
uncomplicated. We start with proving the forward conditional \0
 \left[\frac{\de J}{J} =\alpha ~~\Rightarrow~~ J=\|\langle E_1,E_2 \rangle\|\,C_\theta \right]
\0. By definition we are given that \be \alpha = \de E_1 / E_1 =
\de E_2/E_2 \la{alphaeq}. \end{equation} Consider the vector \0
\langle E_1,E_2 \rangle \0 and its norm \0 h=\|\langle E_1,E_2
\rangle\|= \sqrt{E_{1}^2+E_{2}^2}\, \0. Given simultaneous
increases in enzyme concentration \0 \de E_1 \0 and \0 \de E_2 \0
that obey \equ{alphaeq} we have \0 h+ \de h
=\sqrt{(1+\alpha)^2E_1^2 + (1+\alpha)^2E_2^2} \0. Hence \0 \de h/
h = \alpha \0 and thereby \be \de J/ J = \de h/ h \la{dejdeh}
.\end{equation} In its continuous form equation \equ{dejdeh} can
be set up as a separable differential equation
 such that
\be
(\forall~{E_1}/{E_2}):\left[ \frac {dJ}{J}=\frac{dh}{h} \right].
\la{sepdiff} \end{equation} Integrating both sides of
\equ{sepdiff} we have the solution
\be
(\forall ~E_1/E_2 ):(\exists C_{\theta}): \left[ J= h \,
C_{\theta} \right]. \la{linsol} \end{equation} The backward
conditional \0
 \left[\frac{\de J}{J} =\alpha ~~\Leftarrow~~ J=\|\langle E_1,E_2 \rangle\|\,C_\theta \right]
\0 can be proven by differentiating \equ{linsol} with respect to
\0 h\0.

\0~~~~~~~~~~~~~~~~~~~~~~~~~~~
~~~~~~~~~~~~~~~~~~~~~~~~~~~~~~~~~~~~~~~~~~~~~~~~~~~~~~~~~~\blacksquare{\bf
~Proposition~ \ref{linequiv} }\0\\


 \subsection*{B.3~~Proof of proposition {\ref{theorem2}}}
Consider a case where we posit the existence of a pathway \0
p_1\in\spuJ\0  which obeys the flux summation theorem . In such a
case, using (\ref{afour}) we have;
\be
\bea
 (\forall \theta \mid 0\leq \theta \leq {\pi}/{2}):\\
{ [~[~ h\rightarrow\infty ~]~\ifif~
 [~E_1\rightarrow\infty \andd E_2\rightarrow\infty~]~]}.
 \end{array}
 \end{equation}
Furthermore for the function \0g_{st}\0 defined in \equ{aa37} we
have the following implication:
\be
\bea \left((\exists p_1\in\spuJ) : \left[ (\forall E_1,E_2 \in
\real{1}_{> 0}):\nsum \cij =1 \right]\right):\quantone :\\
\limit{h}{\infty}\,J=\limit{h}{\infty}\, g_{st}[E_1,E_2]=
\limit{h}{\infty}\,h\cdot \cos c_1 \cdot f_{st}[\tan c_1]=
\limit{h}{\infty}\,h\,C_\theta=\infty.
\end{array}
\la{a041}\end{equation} On the other hand consider a
diffusion-limited pathway \0p_2\in\spuDJ\0 where \0J=g[E_1,E_2] \0
as defined by (\ref{e30}). When the system reaches a steady state
as defined by (\ref{sdef3b}), we know by (\ref{stoichiometry})
that \0 \dot{\xi}(t)=0 \0 and \0 0=I_{in}(t)-I_{rev}(t)
-\dot{o}(t)\0. Since \0 \limit{t}{\infty}\dot{o}(t)=J^*\0, using
(\ref{diffusion limited}) we have the steady state condition:
\[
\bea {J^* =D\, u(t)- D \, s_1(t) \mbox{~~~~and}}\\
{J\pm\varepsilon= D\, u(t) -Ds_1(t)}.
\end{array}
\]
Since \0 \dot{u}(t)=0 \0; \be\ \bea {(\forall p_2\in \spuDJ
):(\exists D):(\forall u(0)=c_2):(\forall \varepsilon):}\\ {[~c_3=
D \, c_2~]   \andd [~J\pm\varepsilon= c_3 -Ds_1(t)~]}.
\end{array}
\end{equation}
Recall that \0D\geq0\0
 by
(\ref{diffusion limited})
 and \0s_1(t) \geq 0 \0  by
(\ref{def2}). For any finite \0 0<\varepsilon<\infty \0 we can
define a constant \0 c_4 \0 such that \0 c_4=c_3+\varepsilon \0.
Consequently
\be
J\leq c_4
\end{equation}
where we know that  \0 0<c_4<\infty \0. Hence
\be
\bea (\forall p_2\in \spuDJ ):(\forall (E_1,E_2)\in
\real{1}_{>0}:\\ \limit{h}{\infty}\,J=\limit{h}{\infty}\,
g[E_1,E_2] \leq (c_4<\infty).
\end{array}
\la{a40} \end{equation}
If we now posit a pathway \0 p_3\0 that both obeys the flux
summation theorem and is diffusion-limited, propositions
\equ{a041} and \equ{a40} lead to the proposition:
\be
\bea
 \exists (p_3\in \spuDJ):
 \left(\bea
 {  [~p_3 \in \spuJ~] \andd
\left[\quantone : \nsum \cij =1 \right]} \andd \\
{~~~~~~~~~~~~~~~~~~~~~  [~\infty \leq (c_4<\infty)~]}
\end{array}\right)
\end{array} \la{aaa42}
\end{equation}
which is a contradiction. Therefore
\[
   (\every p \in \spuDJ):
   \left[~~
 \neg \left[\quantone:~ \nsum \cij =1
~\right]  ~~\right].
\]
\\
\0~~~~~~~~~~~~~~~~~~~~~~~~~~~
~~~~~~~~~~~~~~~~~~~~~~~~~~~~~~~~~~~~~~~~~~~~~~~~~~~~\blacksquare{\bf
~proposition~ \ref{theorem2}}\0


\subsection*{B.4~~Proof of proposition {\ref{theorem4}}}

For a pathway \0 p\in \spuVJ \0, consider a case in which enzyme 2
is saturable such that
\be
(\forall E_1\in \real{1}_{> 0}) :~~~\left[~ J \leq E_2 k_{cat(2)}
~~\andd~~ J' \leq E_2 k_{cat(2)}~\right],
\end{equation}
where for any \0\de E_1 \0, \0 J'=J+\de J \0.
 It follows that if \0E_2\0 is held constant,  for any \0\de E_1\0
\be
\frac{\delta J}{\delta E_1}\leq \frac{E_2 k_{cat(2)}-J}{\de E_1}.
\end{equation}
Multiplying both sides by \0 E_1/J \0 we have
\be
\frac{E_1}{J}\,\frac{\delta J}{\delta E_1}\leq \frac{E_1 E_2
k_{cat(2)}-J E_1}{J \de E_1}.
\end{equation}
Hence by \equ{cij} we obtain
\be
C_{1}^{J} \leq ~\frac{E_1 E_2 k_{cat(2)}}{J \de E_1}-
\frac{E_1}{\de E_1}. \la{ineqcij1}
\end{equation}
In a similar fashion if enzyme 1 is saturable and \0E_1\0 held
constant, for any \0\de E_2\0 we have
\be
C_{2}^{J} \leq ~\frac{E_1 E_2 k_{cat(1)}}{J \de E_2}-
\frac{E_2}{\de E_2}.  \la{ineqcij2}
\end{equation}
For each enzyme \0 i \0, consider a measure of saturation \0 Sat_i
\0 such that
\be
Sat_i={J}/{E_i k_{cat(i)}}.
\end{equation}
It follows that for any enzyme \0 i \0,
\be
\limit{Sat_i}{1}{\left(\frac{E_i k_{cat(i)}}{J}\right) = 1 }.
\la{sati}
\end{equation}
Denoting \0 R.H.S.\equ{ineqcij1} \0 as the right hand side of
\equ{ineqcij1}, as enzyme 2 approaches saturation, \equ{sati} and
\equ{ineqcij1} imply that
\be
\limit{Sat_2}{1} \left( R.H.S.\equ{ineqcij1}\right)=
\frac{E_1}{\de E_1}-\frac{E_1}{\de E_1}=0.
\la{satapproach1}\end{equation} Similarly, if enzyme 1 is
approaching saturation, \equ{sati} and \equ{ineqcij2} imply that
\be
\limit{Sat_2}{1} \left( R.H.S.\equ{ineqcij2}\right)=
\frac{E_2}{\de E_2}-\frac{E_2}{\de E_2}=0.
\la{satapproach2}\end{equation} Consequently when both enzyme 1
and enzyme 2 approach saturation, \equ{satapproach1} and
\equ{satapproach2} imply that
\be
(\every p \in \spuVJ):
   \left[~(\forall \alldelta):~
   \limit{Sat_1,Sat_2}{1} \left( {\sum_{i=1}^{2}} \cij \right)=0 ~~\right].
\la{cijsumzero}\end{equation}

 \ \\ \0~~~~~~~~~~~~~~~~~~~~~~~~~~~
~~~~~~~~~~~~~~~~~~~~~~~~~~~~~~~~~~~~~~~~~~~~~~~~~~~~~~\blacksquare{\bf
~proposition~ \ref{theorem4}}\0


\subsection*{B.4~~Proof of proposition {\ref{theorem4_2}}}

For a pathway \0 p\in \spuVJ \0, consider a case in which enzyme 2
is saturable such that
\be
(\forall E_1\in \real{1}_{> 0}) :~~~\left[~ J \leq E_2 k_{cat(2)}
~~\andd~~ J' \leq E_2 k_{cat(2)}~\right], \la{cond452}
\end{equation}
where for any \0\de E_ 2 \0, \0 J'=J+\de J \0.

Given that
\0
Sat_i={J}/{E_i k_{cat(i)}},
\0
if \0 E_1 \0 is held constant, for any \0 \de E_2<0 \0 then
\be
\limit{Sat_2}{1}{\left(\frac{E_2 k_{cat(2)}}{J}\right) = 1 } \mbox{   ~~~ and~~~       }
\limit{Sat_2}{1}{\left(J\right) = E_2 k_{cat(2)} }.
\la{sat2222}
\end{equation}
Given \equ {cond452} and \equ{sat2222}, for domains in which  decreasing \0 E_2 \0 increases saturation  we have:\\
For all intervals \0  a < E_2 < b \0 : \\
If \0\frac{ \partial Sat_2} {E_2} \geq 0 \0 ~~and~~
\0 a \geq 0 \0  ~~and~~
\0 \de E_2<0 \0
~~then:
\be
 \limit{Sat_2}{1}{\left(\de J \right) =\de E_2 k_{cat(2)} }.
\end{equation}
Hence
\be
\limit{Sat_2}{1}{ \left( C_{2}^{J} \right) } =
\limit{Sat_2}{1}{ \left(   \frac {\de J/J } {\de E/E }  \right) } =
\frac {\de E_2  k_{cat(2)}/ k_{cat(2)} } {\de E_2 / E_2 } = 1.
\la{sat3333}
\end{equation}
Similarly if enzyme 1 is saturable we can show that
\be
\limit{Sat_1}{1}{ \left( C_{1}^{J} \right) } =  1.
\la{sat33332}
\end{equation}
Hence
for all intervals \0  a_i < E_i < b_i \0 :

\be
\bea
(\every p \in \spuVJ):\\
\left(
\bea
\left[~ \forall i  : \left(~
  \frac{ \partial Sat_i} {E_i} \geq 0  ~~\andd~~
 a_i \geq 0   ~~\andd~~
 \de E_i<0 ~\right) ~\right] \Rightarrow \\
~~~~~~~~~~~~
 \limit{Sat_1,Sat_2}{1}{\left( C_{1}^{J} + C_{2}^{J} \right) = 2 }
\end{array}
\right)
\end{array}
\end{equation}

\ \\ \0~~~~~~~~~~~~~~~~~~~~~~~~~~~
~~~~~~~~~~~~~~~~~~~~~~~~~~~~~~~~~~~~~~~~~~~~~~~~~~~~~~\blacksquare{\bf
~proposition~ \ref{theorem4_2}}\0


\subsection*{B.5~~Proof of proposition \ref{theorem3}}
  The most straight forward
proof of (\ref{theorem3}) is by direct deduction. To do this we
define a set of notations for representing discrete changes in
steady state flux due to changes in enzyme concentration.

\newcounter{Adefs}
\setcounter{Adefs}{1}


\subsubsection*{Definition B.4a:}\addtocounter{Adefs}{1}
 For a given input \0 u(t) \0 that is constant,
let $J$ be the abbreviated notation for the result of a mapping in
a function \0 g: \mathbb{R}^2 \mapsto \mathbb{R}^1 \0, with the
variables $E_{1}$ and $E_{2}$ as arguments:
\be
\bea (\forall p \in \spuJ):\\ {(\forall u(0)\in \real{1}):
(\forall (\varepsilon >0)\in \real{1}_{\geq 0}): (\exists g:
\fset{2}{1}):}
\\ {[~~J = g[E_{1},E_{2}]  ~~~\ifif ~~~  J=
g_{\varepsilon}[u(0),E_{1},E_{2}] ~~]}
\end{array}
\la{e30} \end{equation}
 The discrete operator
$(~)\stackrel{E_{1}E_{2}}{\longmapsto}$ applied to $J$ is defined
as:
\[J\dc{E_{1}}{E_{2}} \,\equiv~ g[E_{1}+\de E_{1}~,~E_{2}+\de E_{2}] \]
where the symbol ``$\equiv$'' denotes identity by definition and
 for a given
function $g$, the increments in $E_{1}$ and $E_{2}$ are held
constant such that \0 \delta E_{1}=c_1\0, ~\0\delta E_{2} = c_2\0
and \0 c_1,c_2\in \real{1}_{\geq0}\0. Similarly,
\[J\dc{E_{1}}{\,} \,\equiv\, g[E_{1}+\de E_{1}~,~E_{2}] \mbox{~~~~and~~~~}
 J\dc{\,}{E_{2}} \,\equiv\, g[E_{1}~,~E_{2}+\de E_{2}]. \]
 Note that the operator $(~)\dc{(~)}{\,}$ can be applied
 iteratively and is also commutative in its iteration order such
 that
 \[(J\dc{E_{1}}{\,})\dc{\,}{E_{2}}~~=~~J\dc{E_{1}}{E_{2}}~~=~~(J\dc{\,}{E_{2}})\dc{E_{1}}{\,}\]


\subsubsection*{Definition B.5b:}
\addtocounter{Adefs}{1} Let the operator $(~) \overset{(~)}
         {
          \underset{\Delta}{(\longmapsto)}
         }
$ applied to $J\dc{E_{1}}{E_{2}}$ be defined as
\[  J\ddc{E_{1}}{E_{2}} \,\equiv\, J\dc{E_{1}}{E_{2}}-J \] where from
Definition~A1 we get
 $ (J\dc{E_{1}}{E_{2}}-J) = (g[E_{1}+\de E_{1},E_{2}+\de E_{2}] - g[E_{1},E_{2}]) $. Similarly,
\[  J\ddc{E_{1}}{\,} \,\equiv\, J\dc{E_{1}}{\,} - J \mbox{~~~~and~~~~}
   J\ddc{\,}{E_{2}}\,\equiv\, J\dc{E_{2}}{\,} - J. \]
%
\subsubsection*{Proof Argument:}
To prove the relation
\be
\bea \quantone:\\
 \left[~~\forall \alphamove :~~
   \frac{\de J}{J} = \nsum \left(\frac{\de
J}{J}\right)_{i}~~\right]  ~~\Rightarrow~~  \dpj=0, \ena
\la{noepistoprove}
\end{equation}
we start with the left hand side of the relation, which is
equation (\ref{18}) from the Kacser \& Burns derivation. Using
definitions B.5a and B.5b for a two enzyme pathway, equation
(\ref{18}) is equivalent to
\be
\frac{J\ddc{E_{1}}{E_{2}}\,}{J} = \frac{J\ddc{E_{1}}{\,}}{J} +
\frac{J\ddc{\,}{E_{2}}}{J}. \la{a29}\end{equation} By cancelling
the denominator in (\ref{a29}) we proceed with a series of
deductions in which the horizontal symbol ``\0 \Updownarrow \0''
denotes the biconditional relation `` if and only if ''  between
successive lines. The explanation for each biconditional is given
to its right:

\begin{align*}
\quantone :~\forall \alphamove : \\ J\ddc{E_{1}}{E_{2}}\, &=
J\ddc{E_{1}}{\,} + J\ddc{\,}{E_{2}} \\ &\Updownarrow \mbox{~using~
Definitions~ B.5a~ \&~ B.5b} \\
 g[E_{1}+\de E_{1} ~,~ E_{2}+\de E_{2}]-g[E_{1},E_{2}] &=
(g[E_{1}+\de E_{1} ~,~ E_{2}]-g[E_{1},E_{2}]) \\ &~~~~~+(g[E_{1}
~,~ E_{2} + \de E_{2}]-g[E_{1},E_{2}])\\
 &\Updownarrow \mbox{rearrange} \\
 (g[E_{1}+\de E_{1} ~,~ E_{2}+\de
E_{2}]-g[E_{1}+\de E_{1} ~,~ E_{2}]) &-(g[E_{1} ~,~ E_{2} + \de
E_{2}]-g[E_{1},E_{2}])=0 \\ &\Updownarrow \mbox{~using~
Definitions~ B.5a~ \&~ B.5b} \\
 (g[E_{1} ~,~ E_{2} + \de E_{2}]-g[E_{1},E_{2}])\ddc{E_{1}}{\,}~ &=0 \\
 &\Updownarrow \mbox{~using~ Definitions~ B.5a~ \&~ B.5b} \\
 \quantone :~\forall \alphamove : \\
((J)\ddc{\,}{E_{2}})\ddc{E_{1}}{\,} &=0.
\end{align*}
Hence
\be
\bea \quantone:
 ~\forall \alphamove :\\
 ~~\left[
   \frac{\de J}{J} = \nsum \left(\frac{\de
J}{J}\right)_{i}~~\ifif~~((J)\ddc{\,}{E_{2}})\ddc{E_{1}}{\,}
=0~~\right]. \ena \la{funkyequiv}
\end{equation}

\ \\ The proposition \0 ((J)\ddc{\,}{E_{2}})\ddc{E_{1}}{\,} =0 \0
~is operationally equivalent to the sentence in proposition
\ref{theorem2} that proposes the independence of control effects.
 Next it is simple to show equivalence to a continuous version such that
\be
\bea \quantone :\\ \left[~\left[~\forall
\alphamove:~((J)\ddc{\,}{E_{2}})\ddc{E_{1}}{\,} =0~\right]
~~\Rightarrow~~ \dpj=0 ~~\right]. \ena \la{epist
equiv}\end{equation}
 From the third line of our sequence of
biconditional relations we know that the left hand side of
(\ref{epist equiv}) has to comply with the equivalence
\be
\bea
 \quantone :~\forall \alphamove : \\
\left[~~((J)\ddc{\,}{E_{2}})\ddc{E_{1}}{\,} =0  ~~\ifif~~
J\dc{E_1}{E_2}-J\dc{E_1}{\,}-J\dc{\,}{E_2}+J=0 ~~\right]. \ena
\la{discrete}\end{equation}
We also know by the general definition of partial differentiation
that
\[
\spart{J}{E_1}=\limit{\de E_1}{0} \left(\frac
{J\dc{E_1}{\,}-J}{\de E_1} \right) ~~~~\mbox{  and }\]
\be
\dopart{J}{E_1}{E_2}=\limit{\de E_1, \, \de E_2}{0} \left(~ \frac{
J\dc{E_1}{E_2}-J\dc{E_1}{\,}-J\dc{\,}{E_2}+J} {\de E_1 \, \de E_2
} ~\right). \la{partial definition}\end{equation} Substituting
(\ref{discrete}) into (\ref{partial definition}) we get
(\ref{epist equiv}). Using (\ref{epist equiv}) and
\equ{funkyequiv} we get \equ{noepistoprove}.
\\* \0~~~~~~~~~~~~~~~~~~~~~~~~~~~
~~~~~~~~~~~~~~~~~~~~~~~~~~~~~~~~~~~~~~~~~~~~~~~~~~~~~~~~~~\blacksquare{\bf
~proposition~ \ref{theorem3}}\0\\


\subsection*{B.6~~Proof of proposition {\ref{theorem5}}}
From \equ{theorem1} and \equ{linequiv} we know that the summation
theorem implies that
\be
 \forall {\left(\, \alpha = \frac{\de E_1}{E_1}=\frac{\de E_2}{E_2} \,\right)}:  \frac {\de J_{1,2}}{J}=\alpha.
 \la{th51}
\end{equation}
Let
\be
\de E_1 = \alpha E_1  \mbox{~~~and~~~}   \de E_2 = \alpha E_2.
 \la{th52}
\end{equation}
By definition we know that
\be
\bea \quantone : (\forall \alldelta): \\ \nsumtwo \cij = 1
~~\ifif~~  J= \frac{\de J_1}{\de E_1}E_1 + \frac{\de J_2}{\de
E_2}E_2. \la{th53} \ena
\end{equation}
Using \equ{th51} and \equ{th53} we have
\be
\de J_{1,2}= \alpha J =  \frac{\de J_1}{\de E_1}\alpha E_1 +
\frac{\de J_2}{\de E_2}\alpha E_2. \la{th54}
\end{equation}
Using \equ{th54} and \equ{th52} we have
\be
\de J_{1,2}=\de J_{1} + \de J_{2}. \la{th55}
\end{equation}
By \equ{epist equiv}  in the proof for proposition \ref{theorem3}
we know that
\be
\bea \quantone :\\ \left[\left[~\forall \alphamove: ~\de
J_{1,2}=\de J_{1} + \de J_{2}~\right] ~~\Rightarrow~~
\dpj=0\right]. \la{th56} \ena
\end{equation}
\\* \0~~~~~~~~~~~~~~~~~~~~~~~~~~~
~~~~~~~~~~~~~~~~~~~~~~~~~~~~~~~~~~~~~~~~~~~~~~~~~~~~~~~~~~\blacksquare{\bf
~proposition~ \ref{theorem5}}\0\\


\subsection*{B.6~~Proof of proposition {\ref{theorem6}}}
Let \0  J=J[~E_1,E_2~]\0. Given proposition \ref{theorem1} we know
that
\be
J= h C_{\theta} \mbox{~~and~~} J[ 0,0 ]=0. \la{th61}
\end{equation}
By \equ{th61}:
\be
\forall E_1 \in \real{1}_{\geq0}:~~~J[E_1,0]= E_1 C_{E2=0}.
\la{th612}
\end{equation}
Let \[ \de J_{0\rightarrow E1}[~E_1,E_2~] \equiv J[E_1,E_2] -
J[0,E_2].\] Using \equ{th61} and \equ{th612} we have
\be
\bea \forall E_1 \in \real{1}_{\geq 0}:\\ \de J_{0\rightarrow
E1}[~E_1,0~] = J[E_1,0] - J[0,0] = E_1  C_{E2=0}. \ena \la{th62}
\end{equation}
Separately proposition \ref{theorem5} implies that
\be
\bea \forall E_1 \in \real{1}_{\geq 0}:  \exists c_{\Delta}:
\forall E_2 \in \real{1}_{\geq 0}:\\
  ~~~~~~~~~~~~\de J_{0\rightarrow E1}[~E_1,E_2~]= c_{\Delta}.
\ena \la{th63}
\end{equation}
Using \equ{th62} and \equ{th63} we have
\be
\bea \quantone: \\
  ~~~~~~~~~~\de J_{0\rightarrow E1}[~E_1,E_2~]=  E_1  C_{E2=0}.
\ena \la{th64}
\end{equation}
Separately using a procedure similar to deriving \equ{th612} we
have
\be
\forall E_2 \in \real{1}_{\geq 0}:~~~J[~0,E_2]= E_2 C_{E1=0}.
\la{th65}
\end{equation}
Given
\[ J[~E_1,E_2] = J[~0,E_2~] + \de J_{0\rightarrow E1}[~E_1,E_2~] \]
and equations \equ{th64} and \equ{th65} we are left with
\be
J[~E_1,E_2] = E_2 C_{E1=0} + E_1  C_{E2=0}.
\end{equation}
\\* \0~~~~~~~~~~~~~~~~~~~~~~~~~~~
~~~~~~~~~~~~~~~~~~~~~~~~~~~~~~~~~~~~~~~~~~~~~~~~~~~~~~~~~~\blacksquare{\bf
~proposition~ \ref{theorem6}}\0\\


\section*{C.~~Derivations for Numerical Analysis}
\subsection*{Derivations for flux surfaces}
 At steady state the algebraic solution for the
state variables of the equations in
   \equ{odes}  can be obtained using algebraic solving routines. The solution for the seven
internal state variables are in the form of five equations in
which there are no explicit solutions for \0e_1\0 and \0e_2\0 such
that:
\begin{equation}
\bea {s_1=  (e_2 Q (k_2+k_3) k_5 k_7+e_1 k_2 k_4 (Q (k_6+k_7)+e_2
k_6 k_8)) \, \Theta}
            \\ \ \\
  {\underline{es}_{1}=  e_1 k_1 (e_2 Q k_5 k_7+e_1 k_4 (Q (k_6+k_7)+e_2 k_6 k_8)) \, \Theta}
            \\ \ \\
  {s_2=   e_1 k_1 k_3 (Q (k_6+k_7)+e_2 k_6 k_8) \, \Theta}
            \\ \ \\
              {\underline{es}_{2}=  { e_1 e_2 k_1 k_3 k_5 (Q+e_2 k_8) }\,\Theta }
              \\ \ \\
 {s_3=  e_1 e_2 k_1 k_3 k_5 k_7 \, \Theta }
\ena \la{eqsolution} \end{equation} where
\begin{equation}
\Theta = \frac { u D } { e_2 Q (k_2+k_3) k_5 k_7 D+
        e_1 (Q k_2 k_4 (k_6+k_7) D+
              e_2 (Q k_1 k_3 k_5 k_7+k_2 k_4 k_6 k_8 D))}.
\la{eqsolutiontwo}\end{equation}
In addition recall the stoichiometric constraint that
\begin{equation}
\begin{array}{c}
{E_{1}=e_{1}+ \underline {es}_{1} } \\ {E_{2}=e_{2}+ \underline
{es}_{2}.}
\end{array}
\la{stoichotwo} \end{equation}

 Taking the equations \equ{eqsolution}, \equ{eqsolutiontwo} and
 \equ{stoichotwo} we have a system of seven equations. These can be solved by a convergent
 Newton-Raphson
based numerical algorithm for any case where the set of
real-valued parameters \[ \{ D, E_1, E_2, Q,
k_1,k_2,k_3,k_4,k_5,k_6,k_7,k_8 \}
\] is given. The solution for \0 \{ s_1,  s_2,  s_3 ,
e_1,\underline{es}_1,e_2,\underline{es}_2 \} \0 that is arrived at
by numerical approximation can then be used to solve for the
steady state flux \0J\0. The surfaces in Figure \ref{fig2.1} and
Figure \ref{fig2.4} are numerical solutions of \equ{eqsolution},
\equ{eqsolutiontwo} and \equ{stoichotwo}.

 The wild-type kinetic
parameters used in the numerical procedures were \0 k_1=k_5=
4\times 10^7 M^{-1} sec^{-1} \0 , \0 k_2=k_6= 4\times 10^2
sec^{-1}\0, \0 k_3=k_7= 7\times 10^2 sec^{-1} \0 and \0 k_4=k_8=
1\times 10^6 M^{-1} sec^{-1}.\0  The diffusion constants used are
\0 D=Q=3\times 10^1 sec^{-1}\0. For each enzyme \0 K_{eq}=70 \0
and for the wild type \0 k_{cat}=42,000 min^{-1}. \0 Figures
\ref{fig2.5} and \ref{fig2.6} show the steady state solution of
all state variables in the pathway \0 p_{\ref{odes}} \0.

 For simulations of
mutations affecting \0 k_{cat} \0 consider enzyme 1 as an example.
Kinetic mutations have to occur under the thermodynamic constraint
that the equilibrium constant \0 K_{eq}=k_1k_3/k_2k_4 \0 for the
reaction cannot change. Note that \0 V_{max}= k_{cat}\,[E_i]=
k_3\,[E_1] \0. For instance a reduction of \0 k_{cat} \0  to 25\%
of wild type can be modelled by multiplying each of the wild-type
rate constants \0 k_3 \0 and \0 k_4 \0 by 1/4.

\subsection*{Failure of a modified summation conjecture}
A possible counter to the failure of the summation theorem is to
try to include all steps in a pathway, including the non-enzymatic
diffusion steps. Hence a conjecture that
\be
\nsum \cij= C^{J}_{D} + C^{J}_{E1} + C^{J}_{E1} +  C^{J}_{Q}
\la{modifiedc}\end{equation} Here we show the numerical evidence
that such a conjecture does not hold in \0 p_{\ref{odes}} \0. In
any case if such a conjecture had been true, it would not have
been of much practical value since it would mean that for any
pathway deep in metabolism one would have had to take into account
every step starting from the initial diffusion barrier.
A modification of the proof for proposition (\ref{theorem4})
can be used to show the failure of the modified summation
conjecture in regions of saturation. In Figure \ref{fig2.7} we
show the numerical failure of the modified summation conjecture
for a case where \0 \de E_i = 0.02 \mu M\0 and \0 \de D/D=\de Q/Q
= 0.01 \0.



\begin{figure}[pb]
\begin{changemargin}{-.8in}{0.05in}
 \centering
\mbox{\epsfig{file=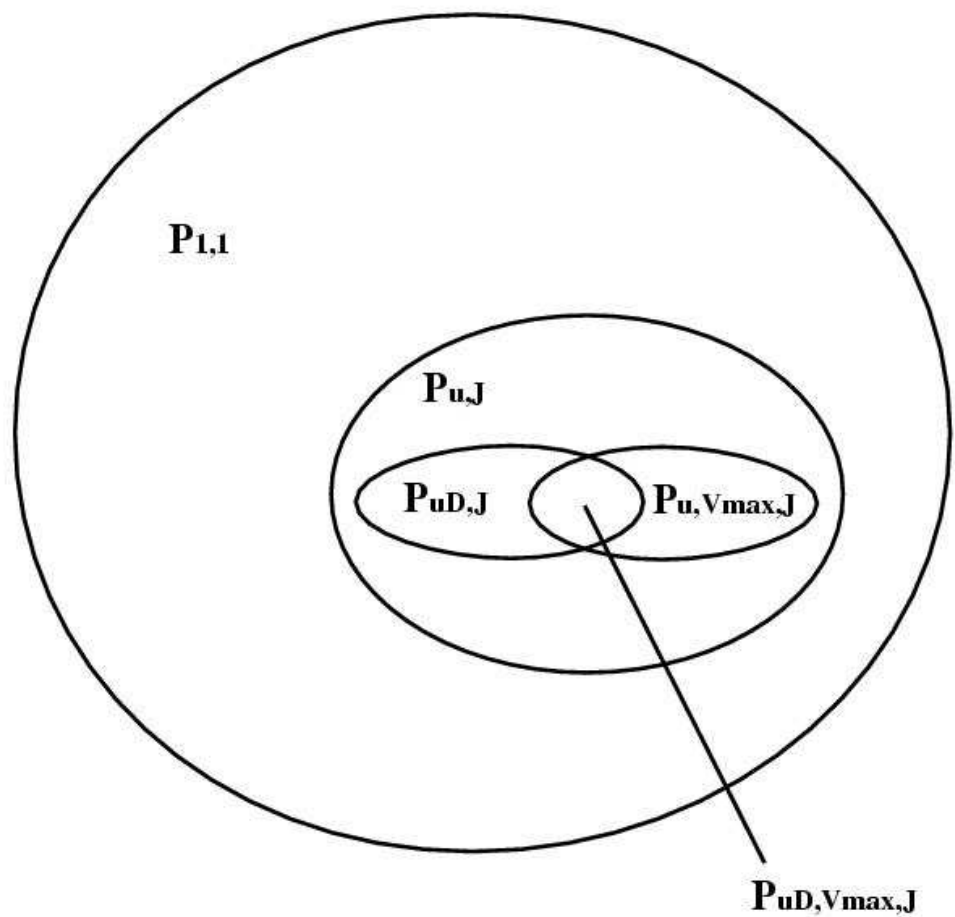,width=17cm}}
\end{changemargin}
\caption{\small \bf The relation between different classes of two
enzyme pathways.
 } \label{fig2.0}
\end{figure}

\begin{figure}[pb]
\begin{changemargin}{-.8in}{0.05in}
 \centering
\mbox{\epsfig{file=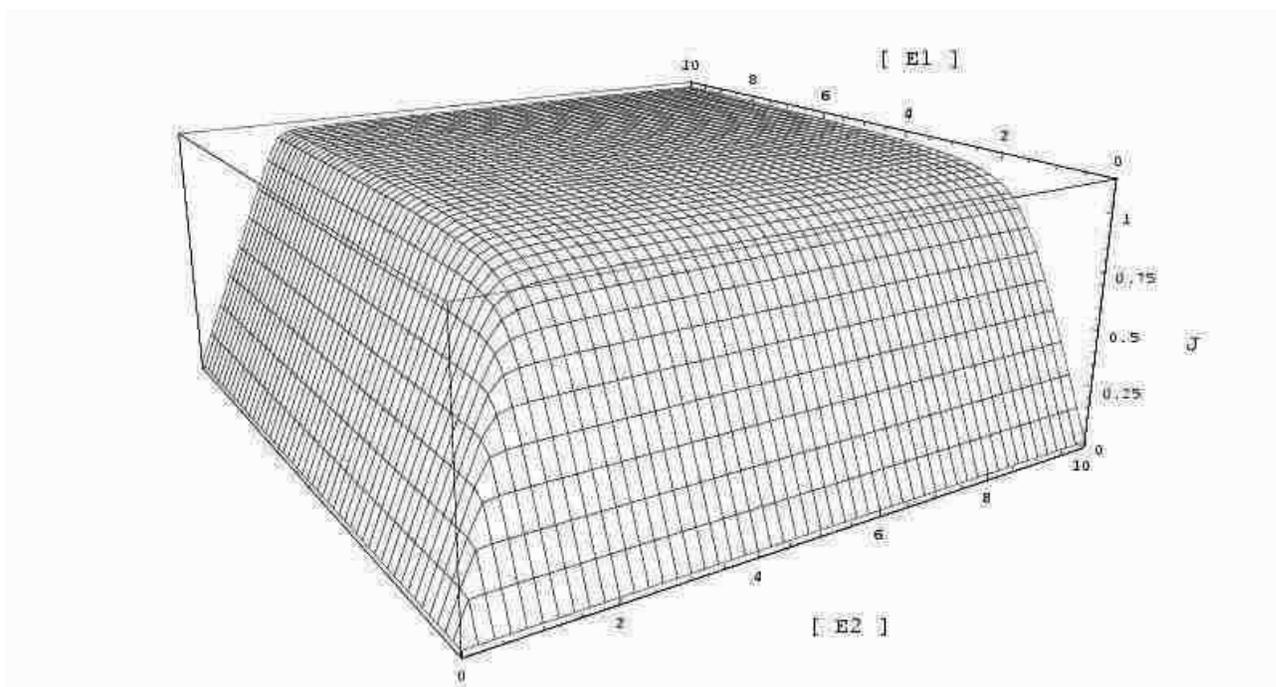,width=17cm}}
\end{changemargin}
\caption{\small \bf Metabolic flux as a function of total enzyme
concentrations. \rm The constant environmental input is  \0 u=
0.75 \0 mM. The catalytic turnover rates are symmetric such that
\0 k_{cat(1)}=k_{cat(2)}=42,000min^{-1}\0. Enzyme concentrations
\0[E_i]\0 in \0\mu M\0. Flux in \0 mM~sec^{-1}\0.
 } \label{fig2.1}
\end{figure}

\begin{figure}[pb]
\begin{changemargin}{-.8in}{0.05in}
 \centering
\mbox{\epsfig{file=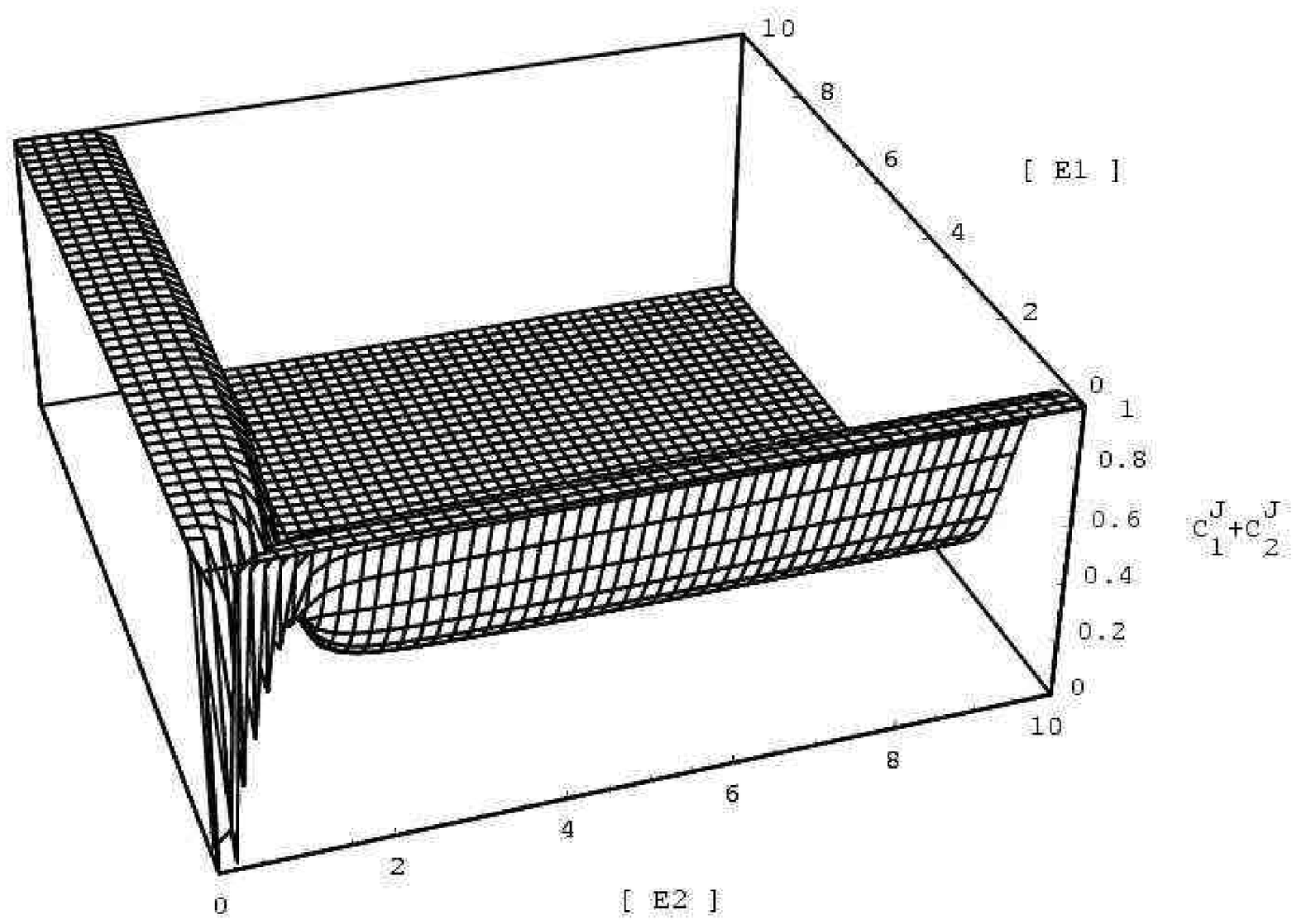,width=17cm}}
\end{changemargin}
\caption{\small \bf Sum of control coefficients as a function of
total enzyme concentrations.\rm ~Enzyme concentrations \0[E_i]\0
in \0\mu M\0. For both enzymes \0 \de E_i = 0.2 \mu M\0.
 } \label{fig2.2}
\end{figure}

\begin{figure}[pb]
\begin{changemargin}{-.8in}{0.05in}
 \centering
\mbox{\epsfig{file=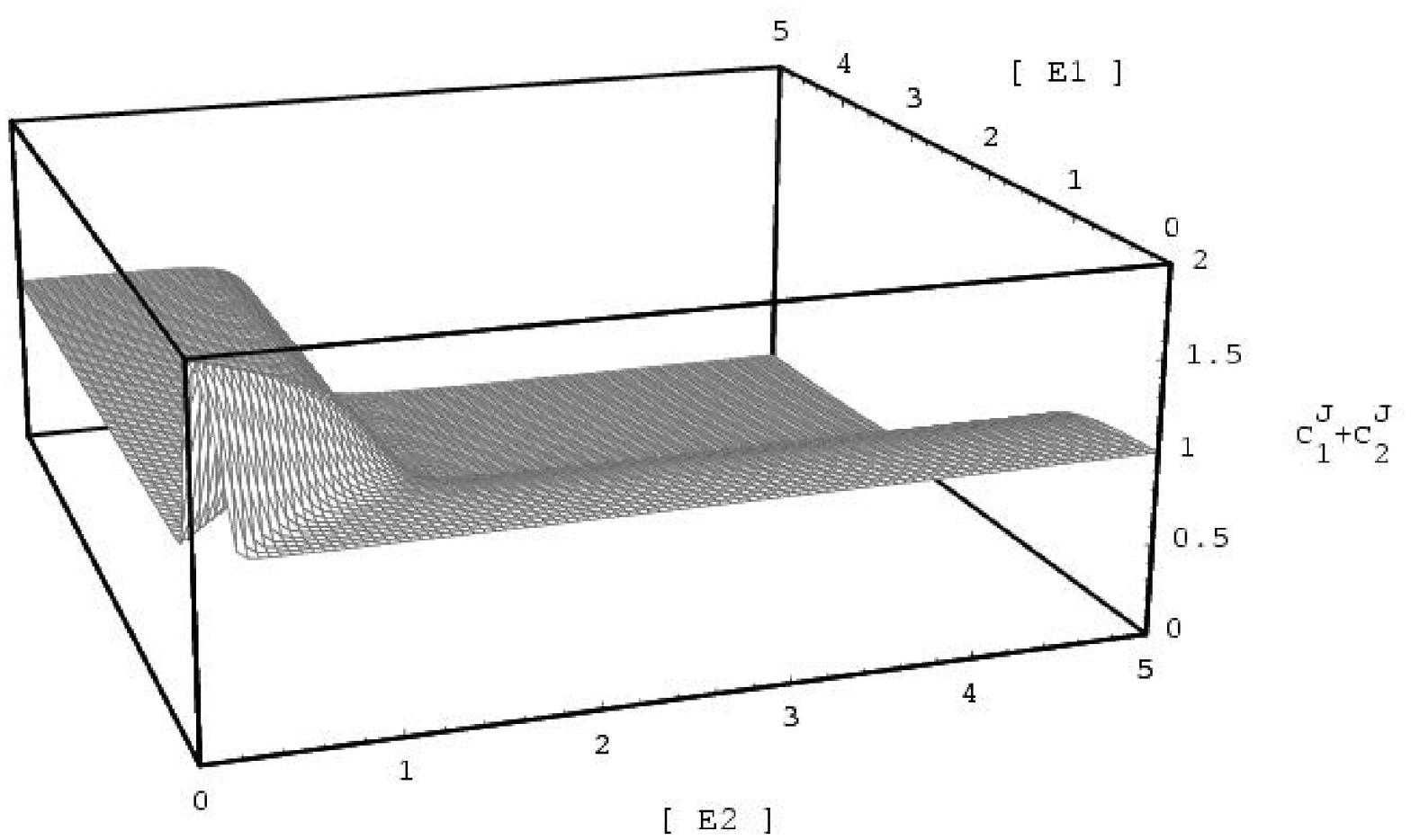,width=17cm}}
\end{changemargin}
\caption{\small \bf Sum of control coefficients as a function of
total enzyme concentrations.\rm ~Enzyme concentrations \0[E_i]\0
in \0\mu M\0. For both enzymes \0 \de E_i = -0.2 \mu M\0.
 } \label{fig2.2.2}
\end{figure}

\begin{figure}[pb]
\begin{changemargin}{-.8in}{0.05in}
 \centering
\mbox{\epsfig{file=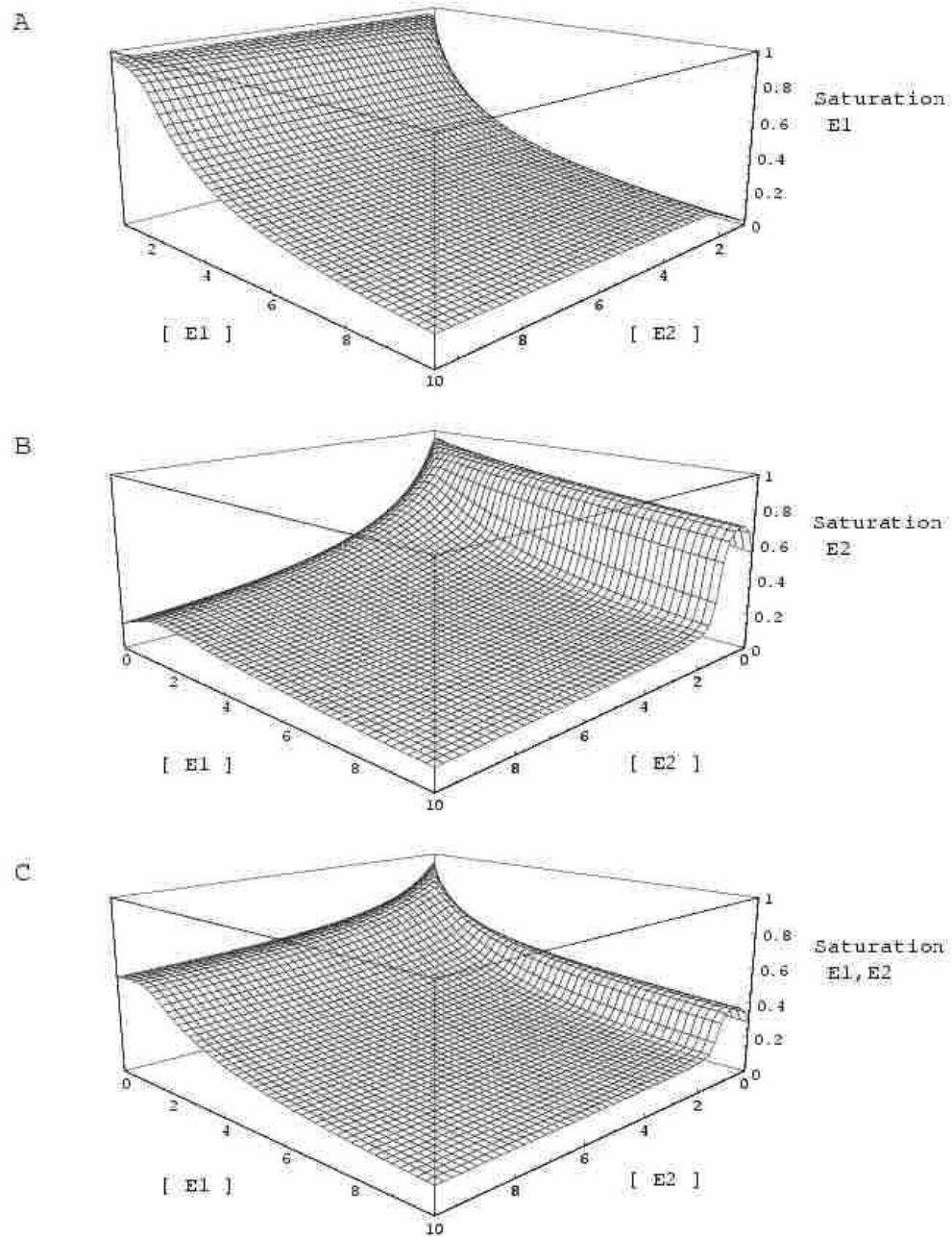,width=17cm}}
\end{changemargin}
\caption{\small \bf Enzyme saturation as a function of total
enzyme concentrations. \rm  Single enzyme saturation is measured
as \0 J/E_i k_{cat(i)}.\0 Double enzyme saturation is measured as
\0 (J/E_1k_3 + J/E_2k_7)/2.\0  Enzyme concentrations \0[E_i]\0 in
\0\mu M\0.
 } \label{fig2.3}
\end{figure}

\begin{figure}[pb]
\begin{changemargin}{-.8in}{0.05in}
 \centering
\mbox{\epsfig{file=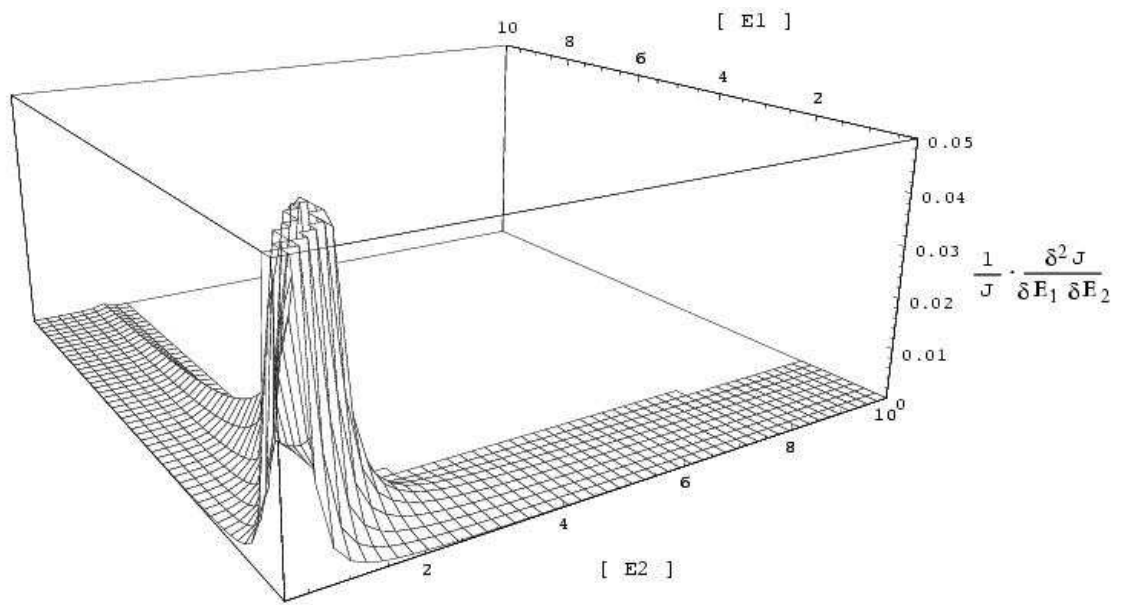,width=17cm}}
\end{changemargin}
\caption{\small \bf Measure of epistatic interaction effects on
flux. \rm Enzyme concentrations \0[E_i]\0 in \0\mu M\0. For both
enzymes \0 \de E_i = 0.02 \mu M\0. } \label{fig2.2-2}
\end{figure}

\begin{figure}[pb]
\begin{changemargin}{-.8in}{0.05in}
 \centering
\mbox{\epsfig{file=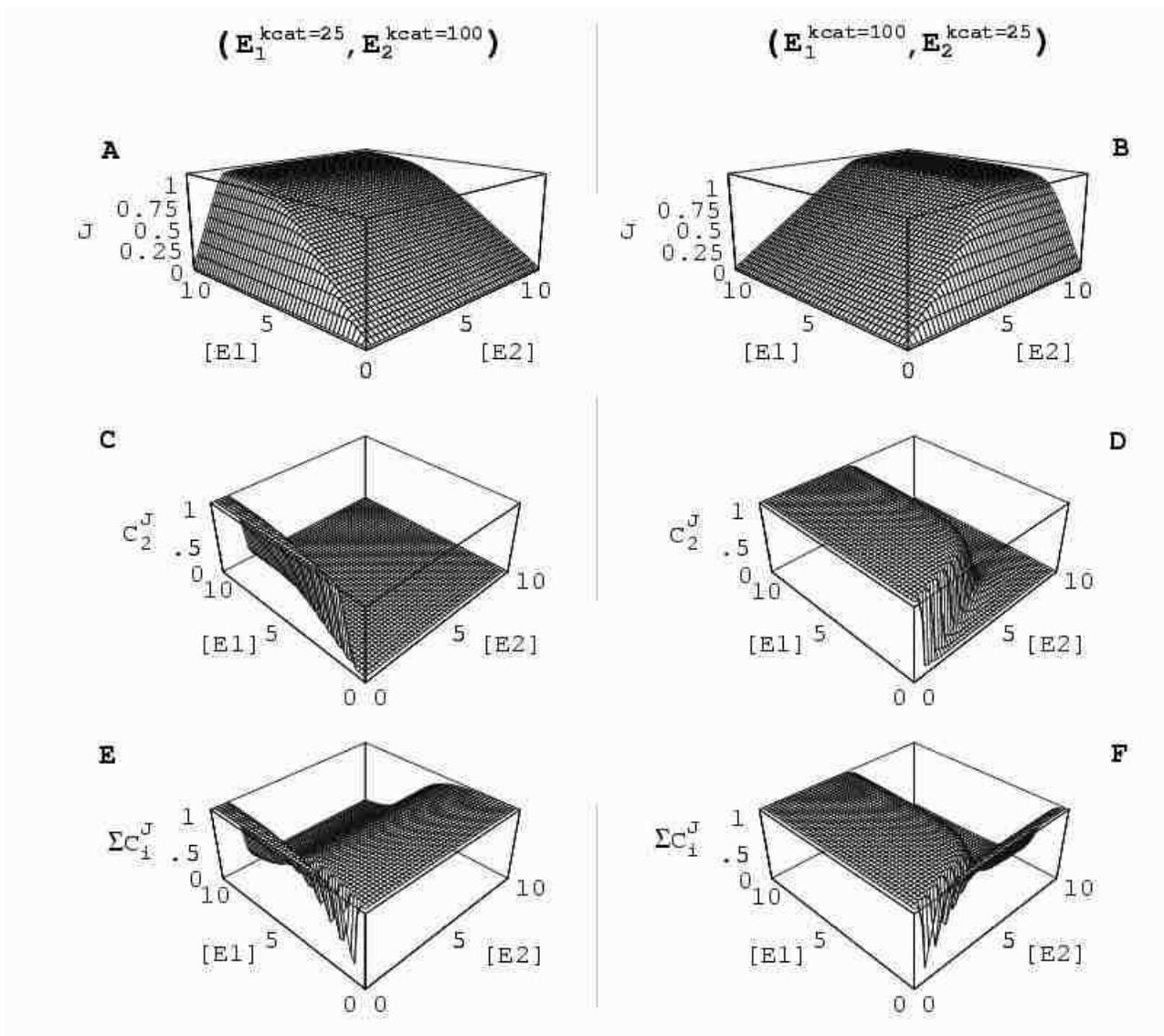,width=17cm}}
\end{changemargin}
\caption{\small \bf Changes in control of a pathway due to
alteration of catalytic turnover rate \0 \bf k_{cat}\0 \rm (\0
k_{cat(i)}=100=42,000min^{-1}\0). \rm Enzyme concentrations
\0[E_i]\0 in \0\mu M\0. For both enzymes \0 \de E_i = 0.02 \mu
M\0.
 } \label{fig2.4}
\end{figure}

\begin{figure}[pb]
\begin{changemargin}{-.8in}{0.05in}
 \centering
\mbox{\epsfig{file=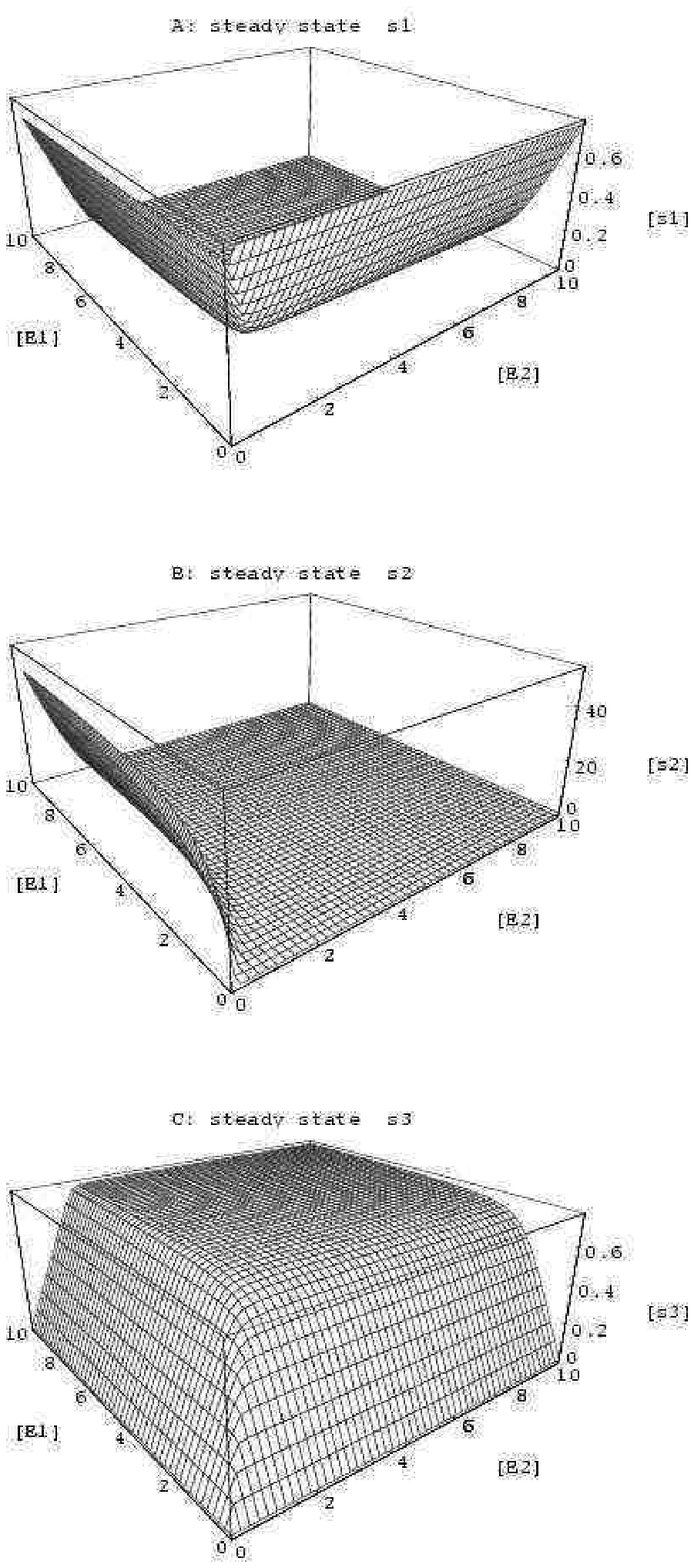,width=17cm}}
\end{changemargin}
\caption{\small \bf Steady state substrate concentrations as a
function of total enzyme concentrations in pathway \0 \bf
p_{\ref{odes}} \0. \rm Horizontal axes: \0\mu M\0 units. Vertical
axes: \0 mM \0 units. (~see Appendix C~).
 } \label{fig2.5}
\end{figure}

\begin{figure}[pb]
\begin{changemargin}{-.8in}{0.05in}
 \centering
\mbox{\epsfig{file=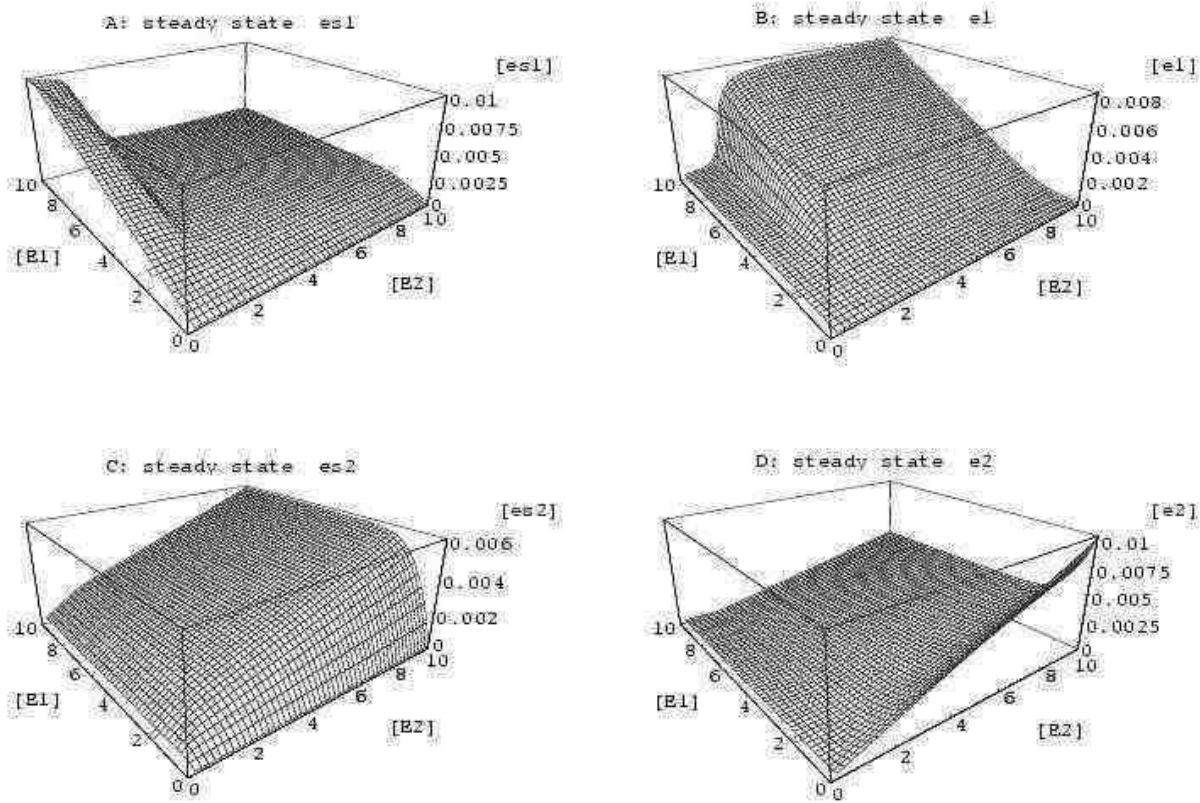,width=17cm}}
\end{changemargin}
\caption{\small \bf Steady state enzyme-substrate and free enzyme
concentrations as a function of total enzyme concentrations in
pathway \0 \bf p_{\ref{odes}} \0. \rm Horizontal axes: \0\mu M\0
units. Vertical axes: \0 mM \0 units (~see Appendix C~).
 } \label{fig2.6}
\end{figure}

\begin{figure}[pb]
\begin{changemargin}{-.5in}{0.05in}
 \centering
\mbox{\epsfig{file=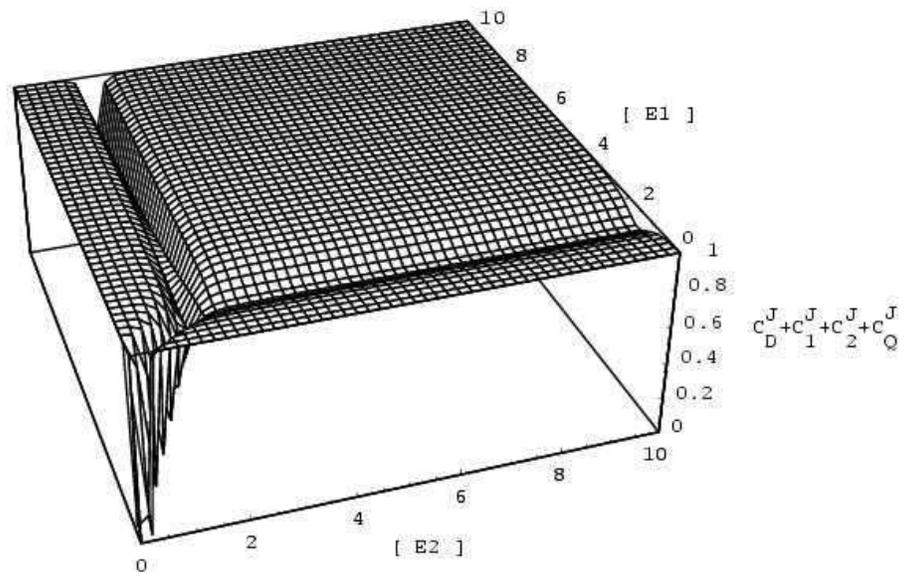,width=17cm}}
\end{changemargin}
\caption{\small \bf Failure of a modified summation conjecture.\rm
Enzyme concentrations \0[E_i]\0 in \0\mu M\0 (~ see Appendix C~).
 } \label{fig2.7}
\end{figure}


\end{document}